\pdfoutput=1

\documentclass[openacc]{rsproca_mod}

\usepackage[export]{adjustbox}
\usepackage{subfig}

\usepackage[margin=0.25in]{geometry}

\usepackage[numbers]{natbib}
\usepackage{subfig}
\usepackage[usenames,dvipsnames]{xcolor}
\usepackage{floatrow}
\usepackage{mwe}
\usepackage{caption}
\usepackage{soul}
\usepackage[colorinlistoftodos,textsize=tiny]{todonotes}
\usepackage{tikz}
\usetikzlibrary{shapes.geometric, arrows}
\usepackage{cancel}
\usepackage{floatrow}

\tikzstyle{Level4} = [rectangle, rounded corners, 
minimum width=2cm, 
minimum height=1cm,
text centered, 
draw=black, 
fill=yellow!90]

\tikzstyle{Level3} = [rectangle, rounded corners, 
minimum width=2cm, 
minimum height=1cm,
text centered, 
draw=black, 
fill=yellow!50]

\tikzstyle{Level2} = [rectangle, rounded corners, 
minimum width=2cm, 
minimum height=1cm,
text centered, 
draw=black, 
fill=yellow!30]

\tikzstyle{Level1} = [rectangle, rounded corners, 
minimum width=2cm, 
minimum height=1cm,
text centered, 
draw=black, 
fill=yellow!10]

\tikzstyle{Level0} = [rectangle, rounded corners, 
minimum width=2cm, 
minimum height=1cm,
text centered, 
draw=black, 
fill=red!0]

\tikzstyle{arrow} = [thick,->,>=stealth]

\usepackage{graphicx}

\newfloatcommand{capbtabbox}{table}[][\FBwidth]

\usepackage{blindtext}
\usepackage{comment}

\begin{document}

\title{Segregation on small rubble bodies due to impact-induced seismic shaking}

\author{
Sohanjit Ghosh$^{1,2}$, Ishan Sharma$^{2,3}$ and Deepak Dhingra$^{4}$}

\address{

$^{1}$Department of Mechanical Engineering, Johns Hopkins University, Baltimore 21210, USA \\
$^{2}$Department of Mechanical Engineering, Indian Institute of Technology, Kanpur 208016, India\\
$^{3}$Department of Space, Planetary \& Astronomical Sciences and Engineering, Indian Institute of Technology, Kanpur 208016, India\\
$^{4}$Department of Earth Sciences, Indian Institute of Technology, Kanpur 208016, India\\
}

\subject{Solar system, Geophysics, Seismology, Impact, Asteroids, Small bodies}


\corres{Sohanjit Ghosh\\
\email{sghosh29@jhu.edu}}

\begin{abstract}
We present a framework to study regolith segregation on rubble-pile asteroids -- self-gravitating granular aggregates -- due to seismic shaking induced by impacts sustained during their lifetimes. We first relate the amplitude and frequency of surface vibrations to the location and severity of an impact, and the rubble body's geometry and bulk properties. For clarity, the body is taken to be an ellipsoid with size and spin close to that of Itokawa, although other asteroids are also easily incorporated. We then model the body's collisional history stochastically given the variability in the impact activity on an asteroid. Finally, we utilize discrete element simulations to investigate the regolith's response to impacts. In these simulations, in any sample collisional history, every time an impact occurs, a bin filled with a grain mixture and located at the region of interest on the asteroid is vibrated at that impact’s associated amplitude and frequency. Utilizing this framework we find that impact-driven seismicity is sufficient to drive size segregation on small rubble-piles, but the segregation quality depends on several aspects, e.g. total impact energy supplied, placement of the region of interest, bulk wave speed and seismic diffusivity.

\end{abstract}





\maketitle
 \section{Introduction}
 \label{sec:intro}

\begin{figure}[b!]
\vspace{-3mm}
\centering\includegraphics[scale=0.8]{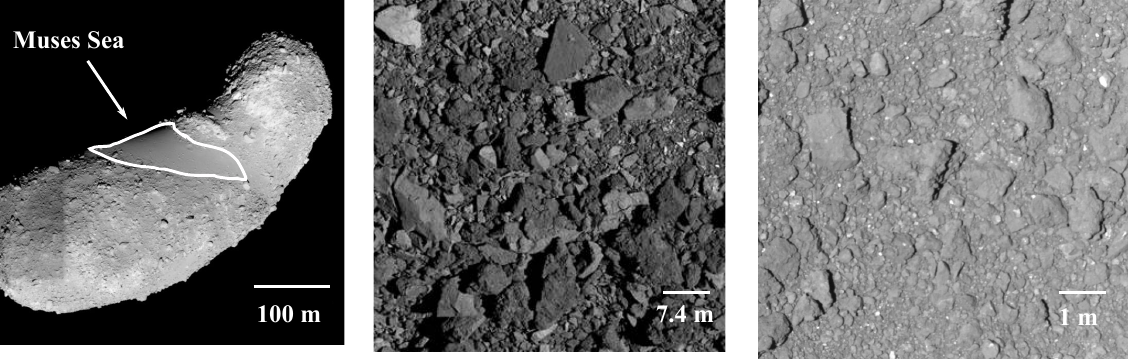}
\caption{Surfaces of \textit{(left)} Itokawa as seen by the Hayabusa spacecraft \textit{(middle)} Bennu near the equatorial region as captured by OSIRIS-REx, and \textit{(right)} Ryugu as observed by Hayabusa2  (credits: ISAS-JAXA/NASA Goddard).}
\label{fig:asteroids}
\end{figure}

Several small bodies in the solar system, e.g. Itokawa, Bennu, and Ryugu, display a prominently granular surface; see Fig.~\ref{fig:asteroids}. High-resolution images at different illumination angles indicate common features between them \cite{fujiwara2006rubble, cambioni2021fine, michikami2019boulder}.  For example, large meter-sized boulders  dominate the surface, against the expectation that thermal fragmentation and micrometeorite impacts should have ground boulders into fine regolith \cite{delbo2014thermal}. This suggests a dynamically active surface, where we may observe segregation of the regolith, even though gravity is several orders of magnitude lower than the Earth's; indeed Itokawa's surface in Fig.~\ref{fig:asteroids}\textit{(left)} shows significant segregation between fines in the Muses Sea and surrounding boulder-rich regions. The main aim here is to investigate if regolith segregation on small rubble asteroids could result from impact-induced seismic shaking.
 
Segregation of different-sized grains in a regolith may be achieved only after the latter is sufficiently mobilized. This, in turn, may occur for a variety of reasons, which we now discuss. The  Yarkovsky-O'Keefe-Radzievskii-Paddack (YORP) effect due to solar radiation torque \cite{bottke2006yarkovsky} may alter the spin of a small body significantly during its lifetime. This may then facilitate large-scale resurfacing, be it through fission or global landslides \cite{scheeres2007effect, graves2018resurfacing}. Tidal encounters with terrestrial planets too may induce global regolith migration \cite{zhang2020tidal}. Impacts are also capable of initiating granular landslides. For example, due to Itokawa's small size, even centimeter-sized impactors may lead to global seismic shaking \cite{miyamoto2007regolith} and subsequent grain motion;  this was investigated in the context of Bennu by \cite{banik2022regolith}.  Finally, non-gravitational forces, such as those due to the interaction of grains and solar wind plasma,  produce electrostatic forces that may sometimes overcome gravity and cause regolith motion  \cite{hartzell2013dynamics,lee1996dust}. 

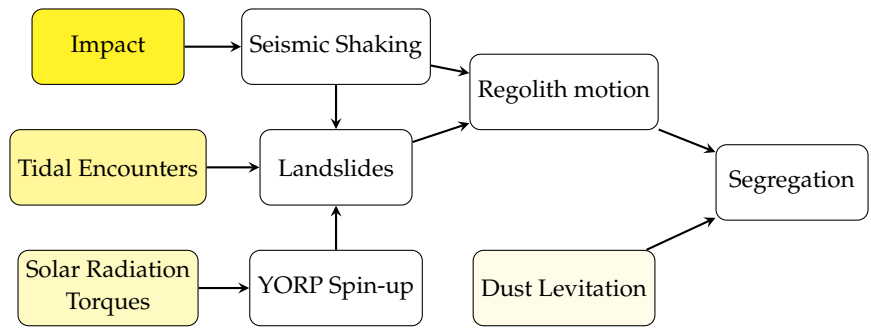
\begin{figure}

\begin{tikzpicture}[node distance=1.5cm]

\node (start) [Level4] {Impact};

\node (in1) [Level0, right of=start,xshift=1.5cm] {Seismic Shaking};


\node (in3) [Level3, below of=start,yshift=-0.1cm] {Tidal Encounters};

\node (in4) [Level2, below of=in3,yshift=-0.1cm,align=center] {Solar Radiation \\ Torques};

\node (in5) [Level0, right of=in3,xshift=1.5cm] {Landslides};

 \node (in6) [Level0, right of=in5,xshift=1.5cm,yshift=1cm] {Regolith motion};

 \node (in7) [Level0, right of=in6,xshift=1.5cm,yshift=-1.2cm] {Segregation};

\node (in8) [Level0, right of=in4,xshift=1.5cm] {YORP Spin-up};

\node (in9) [Level1, right of=in8,xshift=1.5cm] {Dust Levitation};

\draw [arrow] (start) -- (in1);
\draw [arrow] (in1) -- (in6);
\draw [arrow] (in9) -- (in7);
 \draw [arrow] (in6) -- (in7);
\draw [arrow] (in4) -- (in8);
\draw [arrow] (in8) -- (in5);
\draw [arrow] (in3) -- (in5);
 \draw [arrow] (in5) -- (in6);
\draw [arrow] (in1) -- (in5);

\end{tikzpicture}
\caption{Different routes by which regolith on small bodies may be segregated. Primary mechanisms are shaded in proportion to their relative importance in mobilizing regolith on an Itokawa-like body.} 
\label{fig:allprocesses}
\end{figure}

All of the above processes act independently and over different timescales. Moreover, each process affects individual asteroids to varying degrees. Figure~\ref{fig:allprocesses} represents potential routes to granular segregation on an asteroid's surface through a flowchart, in which the primary agents are shaded in proportion of their importance for an asteroid such as Itokawa, and this may change for other bodies. We estimate that impact-induced seismic shaking is most relevant for a body like Itokawa, because tidal effects, YORP, and electrostatic forces are not strong enough to initiate grain motion in this case. We now discuss the reasons for this claim in the following few paragraphs. 

Numerical simulations by \cite{yu2014numerical} of a close encounter with Earth of the near-Earth asteroid (NEA) Apophis, which has size similar to Itokawa, show that Apophis' surface is minimally affected even at its expected closest distance of 20000 miles. On the other hand, Itokawa's closest approach to Earth is much greater at 42 million miles \cite{fujiwara2006rubble}. At the same time, Itokawa is believed to have originated in the main belt, where it would not have been tidally affected by the giant planets. Together this suggests the implausibility of resurfacing on Itokawa by tidal interactions, which do not appear to have been strong enough to mobilize grains. Having said this, we recognize that the possibility and extent of Itokawa's close encounters with Mars have not yet been studied. 















Radiation torque or YORP is a slow process that modifies the spin of asteroids over long periods of time. For Itokawa, \cite{scheeres2007effect}  claim that YORP may have decelerated it from 6.5 h to its current spin state of 12.132 h in 50 -- 90 thousand years. At the same time, \cite{zhang2020tidal} report that surface material failure is generally initiated by YORP-driven spin {\em up}, leading us to contend that YORP is unlikely to have resurfaced a slow spinner like Itokawa.  Separately, we note that YORP is extremely sensitive to topography, with minor changes causing major deviations, even changes in spin direction \cite{banik2022regolith}. Including YORP in any resurfacing model thus requires closely following the surface's evolution.

The possibility of electrostatic interactions of surface grains with charged plasma being able to levitate the finer-sized entities and transport them elsewhere has been reported by \cite{lee1996dust, hartzell2013dynamics}. There is not much knowledge about the extent and distribution of charged grains on Itokawa, or the plasma field in its vicinity. There is also no direct observational evidence of dust levitation on Itokawa. Moreover, it is expected that at leading order the redistribution of grains due to electrostatic interactions will be agnostic to the asteroid's topography, which would then fail to explain the concentration of finer grains at topographic lows, such as Muses C on Itokawa in Fig.~\ref{fig:asteroids} (\textit{left}). Due to these aspects, we neglect the role of electrostatic levitation on segregation. 


In contrast to the above processes that are either infrequent and/or not energetic enough and/or not well understood, impacts on a typical small asteroid are frequent, supply sufficient energy, and impact cratering is a fairly mature subject \cite{melosh1989impact}. Indeed, impacts are an attractive mechanism for resurfacing small bodies whose size allows seismic energy concentration to remain high for a long time after the impact   \cite{cintala1978characteristics}. A typical NEA experiences a large number of impacts during its lifetime, e.g. we estimate in Sec.~\ref{sec:holsapple}(b) that Itokawa experienced about 150,000 small and big impacts over  10 MYr. This is the typical lifetime of an NEA \cite{gladman2000near} which, for Itokawa, includes the conjunction of its head and body \cite{terada2018thermal} and its insertion into a near-Earth orbit from the main belt \cite{michel2013collision}. Finally, due to their low gravity, even gentle seismic activity is able to mobilize surface material on a small asteroid \cite{miyamoto2007regolith}. For these reasons, we highlight impacts as the most important primary mechanism in Fig.~\ref{fig:allprocesses} and focus here on regolith segregation due to impact-induced seismic shaking.

Impacts lead to local shaking and may also induce landslides. On bodies with granular regolith, the former may lead to segregation through the Brazil-nut effect (BNE) \cite{rosato1987brazil}, while kinetic sieving in landslides \cite{vallance2000particle} separates small and large grains.  In terrestrial conditions, BNE has been investigated at length \cite{mobius2001size}, but less so in low-gravity environments, where questions remain about both its presence and efficacy. Employing discrete element (DE) simulations \cite{cundall1979discrete}, BNE was shown by  \cite{tancredi2012granular,ghosh2021granular} to occur even when surface gravity was $10^{-4}\textsl{g}_{_\oplus}$, where $\textsl{g}_{_\oplus}$ is the  gravitational acceleration on the Earth's surface. However, their simulations were greatly simplified in terms of the number and size distribution of grains, and the manner in which energy was supplied to them. 
This work was improved upon by \cite{maurel2016numerical} who surveyed the dependence of  BNE upon the grains' mechanical properties, e.g. inter-grain friction and restitution coefficient.  
All of these studies utilized an arbitrary amplitude and frequency of shaking, unrelated to impact-induced seismic activity. Grain separation due to impact-induced landslides on small bodies has been explored even less, although \cite{gaurav2021granular} probed this possibility in a two-dimensional setting. 

 The above discussion motivates the present work, wherein we first formulate a framework to model impacts over a small rubble-pile asteroid over its lifetime, incorporate the induced seismic activity, and then link this to the segregation of surface regolith due to localized seismic shaking through a BNE-like process.  This framework is then applied to an {\em ellipsoidal} NEA with an aspect ratio similar to that of Itokawa to investigate possible resurfacing outcomes. Although local seismic shaking may in some instances -- depending upon the availability of the seismic energy and the presence of gradients in the effective surface gravity -- release enough regolith to cause a landslide, we will ignore this possibility here. These may be included after appraising the outcomes of the present exercise. 
 
We will describe the collision history of our ellipsoidal rubble pile NEA stochastically. To this end, we take its spin state and age similar to that of Itokawa and follow the methodology of \cite{holsapple2022main}  to determine a size distribution of impacts experienced by the asteroid over its lifetime. From this, we estimate the number of impacts that are substantial enough to initiate global seismic shaking but are not so energetic that they disrupt the asteroid. These impacts are further distinguished between small and large impacts and are then allowed to occur at random time intervals and at random locations over the asteroid's surface, thereby creating several possible collisional histories. 

For each sample collisional history, we model the transmission of seismic energy within the asteroid's volume for every impact in the manner of \cite{richardson2005global}. Seismic properties like quality factor, seismic diffusivity, and seismic frequency that govern this transmission are often not well-determined for rubble-pile bodies. Here, we improve upon the estimates of \cite{richardson2020impact} by accounting for seismic energy transfer through the rubble-pile asteroid's granular and, therefore, highly dissipative interior.  The post-impact seismic energy distribution over the asteroid's surface is then employed to drive local seismic shaking that is studied through DE simulations. In the latter, surface regolith -- modeled as a granular mixture -- is vibrated at an amplitude and frequency that is related to the impact and associated seismic activity; cf. Sec.~\ref{sec:algorithm}(d). When performed for every impact in a sample collisional history we obtain one possible outcome for impact-driven resurfacing of the rubble-pile asteroid. This process is then repeated for several collisional histories. {\color{black} Our simulations indicate that final outcome of impact-induced seismic shaking on the surface of any small body is a complex process that depends on several factors, few of which are addressed in this work. Firstly, we demonstrate that it would suffice to only model the large impacts out of all the impacts capable of causing impact-induced seismic shaking to obtain segregation in a granular mixture. Next, we study how the degree of segregation varies at different locations of a non-spherical target body due to the different amounts of seismic energy received from impacts over its collisional history. We also study the outcome of a granular system subjected to seismic shaking for different initial grain configurations and also different collisional histories and comment on the sensitivity. Finally, in this work, we demonstrate how the seismic properties that govern energy transmission in a small body affect the process of segregation.

}

The rest of the paper is organized as follows. Section~\ref{sec:richardson} models seismic energy transmission from the point of impact to different surface locations of an ellipsoidal NEA. The seismic properties that govern this transmission of energy are discussed in detail in the context of small rubble bodies. Section~\ref{sec:holsapple} discusses the stochastic modeling of an asteroid's impact history.  In Sec.~\ref{sec:algorithm} we couple seismic energy received at a location to DE simulations of vertically vibrated surface grains at that location. Regolith segregation over the NEA's lifetime for several collisional histories is then collated and discussed in Sec.~\ref{sec:results}, before we conclude in Sec. \ref{sec:concl}. 

\section{Impact-induced seismic activity}
\label{sec:richardson}

\begin{figure}[b!]
\begin{center}
\vspace{-8mm}
\subfloat[][]{\includegraphics[scale=1.0]{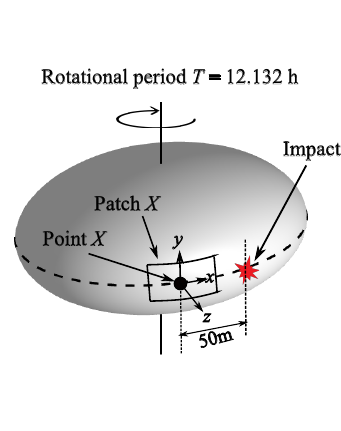}\label{fig:asteroidX}}\hspace*{3em}
 \subfloat[][]{\includegraphics[scale=0.7]{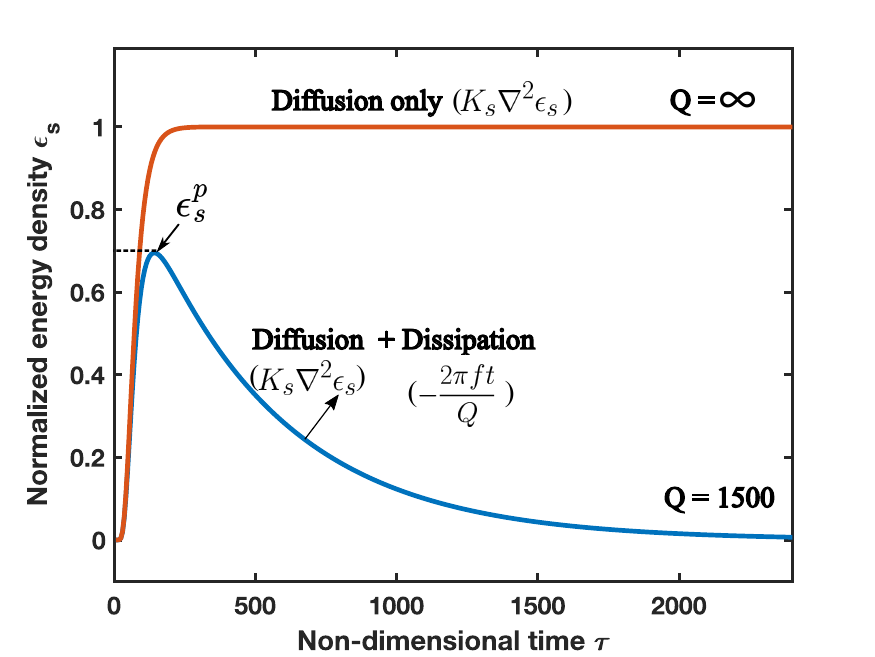}\label{fig:signature} }
\caption{ (a) A schematic of the ellipsoidal NEA that we investigate. The dimensions, density, and spin of the asteroid are  similar to that of Itokawa. A location $X$  at the equator along with a local ``patch" is  indicated. (b) Variation of seismic energy density $\epsilon_s$ with non-dimensional time $\tau$  at point $X$ due to an impact 50m away along the equator as shown in (a) for two values of the quality factor $Q$. The seismic properties employed  are discussed in Sec.~\ref{sec:richardson}\ref{sec:seisprop}.}
\end{center}
\end{figure}

Impacts occur on planetary bodies  frequently. Unlike the Earth, which has an atmosphere, all objects that cross the orbits of small bodies, such as asteroids, reach their surface thereby modifying it significantly over its  lifetime. In this section, we  discuss the distribution of impact energy across the target body and related seismic activity. For this we will extend the methodology of \cite{richardson2005global} to a rubble asteroid, i.e. an asteroid that is comprised of grains held together largely by self-gravity. The asteroid is taken to be a triaxial ellipsoid rotating about its shortest principal axis, as shown in Fig.~\ref{fig:asteroidX}. We limit ourselves to studying the impact-driven resurfacing of an asteroid so that  impactors are assumed to not be large enough to disrupt the asteroid or  alter its shape significantly.



\subsection{Seismic energy transmission}
 \label{sec: seismicenergypropagation}
Out of the total kinetic energy deposited by an impactor, only a small fraction, typically much less than 1\%, called \textit{seismic energy}, is propagated as waves through the body \cite{melosh1989impact}. The energy in these {\em seismic waves} subsequently induces seismic shaking that may mobilize regolith. Following \cite{richardson2005global}, the transmission of seismic energy through the impacted body is modeled as a diffusive process, analogous to heat conduction: 
\begin{equation}
\label{eq:2.1}
\frac{\partial \epsilon_s}{\partial t} = K_s \nabla^2 \epsilon_s - \frac{2\pi f}{Q} \epsilon_s,
\end{equation}
where $\epsilon_s$ is the {\em normalised seismic energy density}, $t$ is time,  $K_s$ is the {\em seismic diffusivity}, $f$ is the {\em seismic frequency}, and $Q$ is the {\em seismic quality factor}. The diffusivity $K_s$ regulates how fast the seismic energy  diffuses through the medium. Further, seismic energy at a given location is dissipated during vibrations initiated there by seismic waves, and this temporal decay is controlled by the quality factor $Q$, assuming that  energy is lost mostly to viscous processes.   This dissipation enters on the right-hand side of \eqref{eq:2.1}. The quality factor is defined \cite{lambeck2005earth} as the inverse of the fraction of energy lost per cycle in a viscously damped harmonic motion: 
\begin{equation}
    \displaystyle \frac{\mbox{Total energy stored in the oscillating body}}{\mbox{Energy lost per cycle}}=\frac{Q}{2\pi}=\frac{E}{\delta E} \implies \delta E = \frac{E}{2\pi Q}.
    \label{eq:2.7}
\end{equation}

Solving \eqref{eq:2.1} over a complex three-dimensional shape, even a triaxial ellipsoid, is analytically cumbersome. Thus, we replace the impacted asteroid by its best-fit cuboid with dimensions $L \times W \times  H$, and impose the following initial and boundary conditions:
\begin{subequations}
\begin{flalign}
\label{eq:2.2}
&&\epsilon_s (x_0,y_0,z_0,t = 0) &= \delta(x-x_0,y-y_0,z-z_0)& \\
\label{eq:2.3}
&\mbox{and, for all $t$,}  &  \frac{\partial \epsilon_s}{\partial x}=0 \mbox{ at $x = 0, L$};  &\ \  
\frac{\partial \epsilon_s}{\partial y}=0 \mbox{ at $y = 0, W$}; \ \ \frac{\partial \epsilon_s}{\partial z}=0 \mbox{ at $z = 0, H$};  &
\end{flalign} 
\end{subequations}
where the impact is taken to occur at the location ($x_0,y_0,z_0$) and $\delta(\cdot)$ represents the delta distribution. The initial condition in (\ref{eq:2.2}) assumes that the seismic energy imparted to the target body may be modeled as a delta distribution, because the target body is much larger than the impactor, as mentioned above. The boundary condition (\ref{eq:2.3}) reflects the fact that seismic energy cannot flow across the body's free surface. The solution for the seismic energy density may now be obtained in closed form as  
\begin{multline}
\label{eq:2.4}
\epsilon_s(x,y,z,t) = e^{-2 \pi (ft/Q)}\Bigg(1+2\sum_{n=1}^{\infty} \cos{\frac{n \pi x_o}{L}}\cos{\frac{n \pi x}{L} e^{-K_{s} n^2 \pi^2 t /L^2}}\Bigg) \Bigg(1+\dots \\
\dots 2\sum_{n=1}^{\infty} \cos{\frac{n \pi y_o}{W}}\cos{\frac{n \pi y}{W} e^{-K_{s} n^2 \pi^2 t /W^2}}\Bigg)
 \Bigg(1+2\sum_{n=1}^{\infty} \cos{\frac{n \pi z_o}{H}}\cos{\frac{n \pi z}{H} e^{-K_{s} n^2 \pi^2 t /H^2}}\Bigg),
\end{multline}
where $n$ may be identified as  the {\em wave number}. 

Although approximate, the energy distribution in the best-fit cuboid of the asteroid provides a reasonable first estimate of how seismic energy is dispersed through the asteroid's volume. Figure \ref{fig:signature} reports the temporal variation in the seismic energy received at the equatorial point $X$ in Fig.~\ref{fig:asteroidX} when an impact occurs 50m to its East. In Fig.~\ref{fig:asteroidX} we employ a non-dimensional time $\tau$ obtained by scaling time by the  time it takes for a $P$-wave to travel 50m in  basalt; this turns out to be 0.0167s, as the $P$-wave speed in basalt is 3 km-s$^{-1}$. From Fig.~\ref{fig:signature}, we observe that there is an initial rise or {\em build-up}  of  seismic energy when the seismic wave reaches point $X$. This is followed by a gradual decay due to both local dissipation and diffusion away. The efficiency of the latter two processes may be gauged from Fig.~\ref{fig:signature} by comparing the evolution for $\epsilon_s$ when $Q=1500$ with when  $Q\to\infty$, at which limit there is no viscous dissipation. 



We now relate the manner in which the surface shakes in response to the seismic energy available at that location. Local surface vibrations will vary in frequency and amplitude depending upon  material properties such as stiffness and dissipation, as well as the available seismic energy and its spectral profile. This is a complex dependency that is compounded by the lack of detailed information about the body's properties. As a first step, we 
assume that the energy received through seismic waves at any surface location of the asteroid excites local harmonic vibrations at an amplitude $A$ and  {\em seismic frequency}  $f = \omega/2\pi$. The energy density $\epsilon_d$ stored in such a vibration is 
\begin{equation}
 \label{eq:2.5}
\epsilon_d = \frac{1}{2}\rho_a \omega^2 A^2 =\frac{\rho_a a_{\max}^2}{8 \pi^2 f^2},
\end{equation}
where  $\rho_a$ is the local surface density and $a_{\max} = A\omega^2$  is the maximum vibrational acceleration. Next, with $\epsilon_s$ given by \eqref{eq:2.4}, we define $\epsilon_s^p= \max_{t \in (0,\infty)}\epsilon_s(x,y,z,t)$ 
 as the maximum energy density experienced post-impact at a given surface location $(x,y,z)$; this is indicated in Fig.~\ref{fig:signature}.
 Assuming that most of the  available seismic energy is transferred to surface vibrations,  we equate $\epsilon^p_s$  with $\epsilon_d$ in \eqref{eq:2.5} to obtain $a_{\max}$ at a particular $(x,y,z)$. This is then non-dimensionalized by the 
local surface gravity $\textsl{g}_a$ to yield the non-dimensional {\em peak surface acceleration}

\begin{equation}
    \displaystyle \Gamma =  \frac{a_{\max}}{\textsl{g}_a} =\frac{A\omega^2}{\textsl{g}_a}.
    \label{eq:2.6}
\end{equation}
Note that $\Gamma$ varies spatially because  both $\textsl{g}_a$ and, potentially, $\omega$ and $A$ vary over the asteroid's surface. The  parameter $\Gamma$  characterizes the intensity of the vibration, or {\em seismic shaking}, at point $X$, and will be utilized extensively. 

\subsection{Selection of seismic parameters}
\label{sec:seisprop}
\begin{figure}
\begin{floatrow}

\capbtabbox{%

	\begin{tabular}{lr} 
		\hline
		Property & Value\\
		\hline
Impact velocity $v_i\, (\text{km-s}^{-1})$ &  10 \\
Quality factor $Q$ (dimensionless) & 1500\\
Seismic diffusivity $K_s$ (km$^2$ s$^{-1}$) & 0.002\\
Seismic efficiency $\eta$  (dimensionless)  &  $10^ {-7}$\\
P-wave velocity $v_p$ (m s$^{-1}$) & 8 \\
Seismic frequency $f$ (Hz) & 0.0197\\
Estimated Young's modulus $Y$ (kPa) & 80\\
Bulk density $\rho$ (kg m$^{3}$) & 3200\\ 

		\hline
	\end{tabular}
}{%
  \caption{Seismic properties that we employ.}%
\label{tab:properties}}
\ffigbox{%
\includegraphics[scale=1]{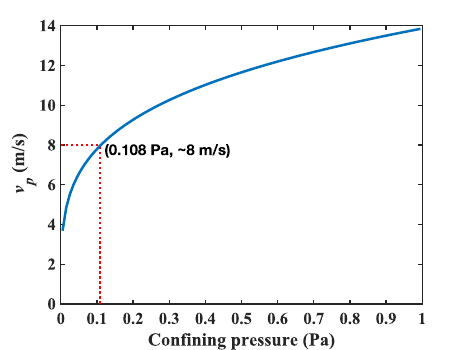} 
}{%
  \caption{Variation of P-wave velocity $v_p$ with confining pressure. }%
\label{fig:pwave} 
}
\end{floatrow}
\end{figure}

The transport of the seismic energy $\epsilon_s$ through the body depends upon its seismic properties; cf. \eqref{eq:2.4}.   Table \ref{tab:properties} summarizes the  seismic properties that we employ for a rubble-pile asteroid. We now discuss these choices in detail.

We first consider the  impact velocity $v_i$. Typical velocities of near-Earth objects (NEOs) are about 10-15 km-s$^{-1}$, which is much higher than the 5 km-s$^{-1}$ of main-belt objects (MBOs) \cite{bottke1993collision}. At the same time, NEAs have their aphelion in the main belt due to their highly eccentric orbits. Impacts between an NEO and an MBO would therefore be at about 10 km-s$^{-1}$, assuming orbital motion in the same direction. Thus, we take  $v_i =  \text{10 km-s}^{-1}$ to  represent impacts  on our Itokawa-like asteroid.



The last term in \eqref{eq:2.1} contains the quality factor $Q$ defined in \eqref{eq:2.7}, for which we now identify an appropriate value. The size-frequency distribution (SFD) of the craters on Itokawa's surface was modeled by \cite{richardson2020impact} through {\em Small Body Crater Terrain Evolution Model} (SBCTEM) simulations. In these simulations, impactors struck a monolithic bedrock layer stochastically, and the evolution of the terrain was observed while varying its $Q$ from 1000 to 2500. The simulations that best fit Itokawa's crater count and size distribution corresponded to $Q = 1500$. We employ this value for $Q$ in our work.

The diffusion rate of seismic energy is governed by the diffusivity $K_s$, with a low $K_s$ indicating a less well-connected body. To estimate $K_s$ we again consider the SBCTEM simulations of \cite{richardson2020impact}. Varying $K_s$ from $0.001 - 0.250$ km$^2$s$^{-1}$ it was found that $K_s$ = 0.002 km$^2$-s$^{-1}$ matched Itokawa's crater distribution best, and we will thus employ this value for $K_s$. A low value for $K_s$ is compatible with our supposition that Itokawa's interior is granular. 

We now verify the consistency of the above estimate of $K_s$ for a granular body. Following \cite{richardson2005global}, and in analogy with three-dimensional thermal diffusion \cite{jeans1982introduction}, we may define 
\begin{equation}
    \displaystyle K_s = v_p l_s/3, 
    \label{eq:2.8}
\end{equation}
where  $v_p$ is the mean seismic velocity and $l_s$ is a `mean free path', in that it is the average distance over which the seismic energy reduces to $e^{-1}$ of its initial value \cite{dainty1974seismic,toksoz1974structure}; $l_s$ is taken to be the average asteroid diameter. The low value of  $K_s= \text{0.002 km$^2$-s$^{-1}$}$ estimated above leads from \eqref{eq:2.8} to  a $P$-wave speed of about 20 m-s$^{-1}$. Such a low value may not be unexpected for gently held granular aggregates, like rubble asteroids. Indeed, we recall \cite{lowrie2020fundamentals} that in an isotropic and homogeneous solid the $P$-wave speed $v_p$ is given by 
\begin{equation}
    \displaystyle v_p = \sqrt{Y\Big/\rho}\, , 
    \label{eq:2.9}
\end{equation}
where  $Y$ is the solid's Young's modulus and $\rho$ is its density.   A porous-granular medium \cite{hardin1963elastic,latham1970passive}, such as our rubble asteroid, would have a low effective elastic modulus compared to a coherent medium due to the presence of internal voids. This would suggest a $P$-wave speed similar in magnitude to the low value obtained above from \eqref{eq:2.8}. However, $P$-wave speeds in self-gravitating rubble-piles are not known. Consequently,  lunar \cite{kovach1973velocity} and Martian values \cite{xu2022modeling,brinkman2022situ} are sometimes adopted, e.g. \cite{quillen2019impact, murdoch2017probing}. Others, e.g.  \cite{richardson2005global, richardson2020impact}, utilize the seismic frequency $f$ for small bodies from seismic experiments conducted on the Earth and the Moon and relate it to $v_p$, as is done further below. Here, we find the $P$-wave velocity in a granular aggregate from a consideration of their response as discrete media.  


 We saw in \eqref{eq:2.9} that the $P$-wave speed in a linear elastic material is given in terms of its bulk stiffness.  The effective stiffness at any point within a granular aggregate depends upon the local confining pressure \cite{murdoch2017probing, goddard1990nonlinear}.   In a geophysical context, the  confining pressure at a given depth may be estimated as the lithostatic pressure  of the overburden. Thus, the confining pressure at a depth $d$ below the surface is $\rho \textsl{g}d$, where $\rho$ is the  bulk density and $\textsl{g}$ is the  surface gravitational acceleration, both  averaged over the depth $d$. In rubble asteroids, due to their small size, gravity is low, and the confining pressure is of the order of a few pascals. For example, considering the density of chondrite to be 3200 kg-m$^{-3}$,  Itokawa's bulk porosity as 40\% \cite{kikwaya2011chondrite} and its averaged surface gravitational acceleration to be $8.5 \times 10^{-5}$ \cite{scheeres2006actual}, the confining pressure at a depth of 1m on Itokawa may be estimated as 
\begin{equation}
p_c \approx  \rho gd = ( 0.4 \times 3200 )\times 8.5 \times 10^{-5} \times 1 = 0.108 \, \mbox{Pa}.
    \label{eq:3.22}
\end{equation}
We now estimate wave speeds in granular media at such low confining pressures. 


Compression waves in randomly packed dry Ottawa sand were experimentally studied at low confining pressures by \cite{hardin1963elastic}. Fitting their experimental data provides an empirical relation for the elastic modulus $Y$:
\begin{equation}
Y = B {(e_c-e)^2/(1+e) }p^{1/2}, 
    \label{eq:2.12}
\end{equation}
where $e$ is the void ratio, and $e_c $ and $ B$ are constants that depend upon the grain type.  The above fit indicates that  $Y$ varies as $p^{1/2}$, so that from \eqref{eq:2.9} we find that the $P$-wave speed $v_p \propto p^{1/4}$. This matches the theoretical estimate of \cite{goddard1990nonlinear} and is consistent with the  $v_p\propto p^{0.19}$ obtained by \cite{jia2021elastic} through DE simulations. We note that we employ results obtained for Ottawa sand because it is siliceous and has a density comparable to chondritic grains found in the regolith of asteroids such as Itokawa, Ryugu and  Bennu \cite{terada2018thermal}.


 Now the experiments of \cite{hardin1963elastic} were performed at confining pressures of the order of kilopascals which, while low in a terrestrial context, is still much higher than those encountered on small bodies. Nevertheless, we extrapolate \eqref{eq:2.12} and utilize the  $p^{1/4}$ variation of the $P$-wave speed with pressure to estimate $v_p$ in a typical rubble asteroid. This is done in Fig.~\ref{fig:pwave} which reports the $P$-wave speed as a function of the confining pressure. We find that $v_p$ is about $8\,\text{m-s}^{-1}$ for the value of confining pressure such as that found on a rubble pile asteroid like Itokawa. This is comparable to the value of $20\,\text{m-s}^{-1}$ estimated from \eqref{eq:2.8} with $K_s= \text{0.002 km$^2$-s$^{-1}$}$. At the same time, utilizing the low $P$-wave speed of $8\,\text{m-s}^{-1}$ in  \eqref{eq:2.9}, yields a very small effective elastic modulus $Y$ of around 80 kPa for Itokawa. This estimate is comparable in scale with that of \cite{richardson2020impact}, whose SBCTEM simulations predicted a value  between 10-20 kPa. Crucially, both our and \cite{richardson2020impact}'s assessments of $Y$ are several orders of magnitude lower than that of the lunar regolith, which is  reported to be of the order of MPa  \citep{kovach1973velocity, toksoz1974structure}. 



As already mentioned,  not all of the impactor's kinetic energy is made available as seismic energy.  Energy is consumed in mobilizing ejecta and in heating and plastically deforming both target and impactor bodies \cite{richardson2005global}. This is characterized through a seismic efficiency $\eta$, which is the fraction of the impact energy that survives as seismic energy $E_i$; thus, 
\begin{equation}
 \label{eq:2.10}
E_i =  \eta E_k = \eta \frac{1}{2} m_i v_{i}^2,
\end{equation}
where  $E_k, m_i$, and $v_i$ are, respectively,  the kinetic energy, mass, and velocity of the impactor. In their SBCTEM simulations, \cite{richardson2020impact} varied $\eta$ from $1 \times 10^{-8} - 1 \times 10^{-6}$ and the simulations that best reproduced the SFD of craters on Itokawa were for  $\eta = 1 \times 10^ {-7}$. This, therefore, is the value of $\eta$ that we will employ. 

The local seismic frequency $f$ occurring in \eqref{eq:2.5} and \eqref{eq:2.6} is another poorly constrained parameter. This, in general, will depend upon the asteroid's material parameters, especially in the vicinity of the surface location of interest. These details are not available at present. We proceed here as follows. We first note that we may ignore energy transfer by surface waves because the local elastic modulus of a granular aggregate vanishes as we approach its surface where the confining pressure drops to zero. At the same time, shear waves are slower than compressional waves. Thus, we may limit ourselves to energy propagation through $P$-waves. Next, appealing to the small size of rubble asteroids, we assume that following an impact the asteroid shakes as a whole, primarily in its {\em first}   mode of vibration, which has frequency $f_0$; thus, the local seismic frequency $f = f_0$.  We now estimate $f_0$ for our rubble-pile asteroid. 

We ignore the asteroid's ellipsoidal shape and consider the vibration of a sphere of equivalent size in its first mode -- also called the {\em breathing} mode -- in which material moves radially. The expectation here is then that impacts preferentially excite radial displacements. The frequency of the breathing mode for a homogeneous linear-elastic sphere with Poisson's ratio $\nu = 1/4$  may be expressed as  \cite{quillen2019impact}:
\begin{equation}
\begin{split}
    f_0 = 0.41 f_{char},
\end{split}
    \label{eq:2.13}
\end{equation}
where $f_{char}$ is the sphere's characteristic frequency, defined as the inverse of the time taken by the $P$-wave to travel across a sphere of radius $R_s$, so that $f_{char}=v_p/R_{s}$.  We will set $R_s = 162$m, which is comparable to the mean radius of Itokawa \cite{fujiwara2006rubble} but, rather than employing $v_p$ in linear-elastic materials, we now employ the value $v_p = 8\,\text{m-s}^{-1}$ obtained above for a granular aggregate.  Once $f_{char}$ is known, we find $f_0$ from \eqref{eq:2.13} and, hence, the seismic shaking frequency $f$ and circular frequency $\omega$ of the rubble asteroid:

\begin{equation}
	 f = f_0 = 0.0197 \mbox{ Hz} \implies \omega = 2\pi f = 0.124 \mbox{ rad-s$^{-1}$}. 
    \label{eq:2.14}
\end{equation}

We now have  estimates of the various parameters governing seismic shaking, and these are collated in table~\ref{tab:properties}. We next describe how we model the collisional history of an NEA such as Itokawa. 

\section{Collisional history}
\label{sec:holsapple}
To investigate texturing at patch $X$ in Fig.~\ref{fig:asteroidX} we need to quantify the total energy received there due to impacts experienced by the asteroid during its lifetime  which, following \cite{gladman2000near}, we take here to  be  around 10 MYr for NEAs. For this, we first need to estimate the number of impacts that the asteroid  may have experienced as a function of the impactor diameter. To this end, we discretize the range of impactor sizes and segregate the impactor population into groups or \lq bins\rq. We then follow the stochastic methodology of \cite{holsapple2022main} to count the number of impacts,  assuming  that the impactor population remains steady with time. We acknowledge that the estimates here will vary for individual  objects, as each asteroid has a different shape and size, and a unique dynamical evolution and collisional history.

We are interested in only those impacts that are  energetic enough to cause resurfacing through global seismic shaking but do not lead to catastrophic disruption. The first step then is to identify the appropriate limits on the impactor sizes. Whenever required, we will take the seismic parameters listed in table~\ref{tab:properties}.

\subsection{Limits on  impactor size}
\label{sec:limits}

Impactors range from microscopic dust particles to other asteroids.  To estimate the minimum impactor diameter $D_{i,\min}$  that may initiate global seismic shaking, we set $\Gamma = 1$ in \eqref{eq:2.6} at the location farthest  from the point of impact. For a spherical asteroid, this is located diametrically opposite to the impact point. In other words, an impactor of diameter $D_{i, \min}$ will initiate  accelerations equal to the gravitational acceleration at the asteroid's surface location most distant from where the  impact took place. Thus, for impactors with sizes greater than $D_{i, \min}$ we expect surface motion everywhere on the asteroid. Inserting $\Gamma=1$ in \eqref{eq:2.6} and employing the estimate for the frequency $\omega$ and the amplitude $A$ from, respectively,  \eqref{eq:2.14} and  \eqref{eq:2.5}, and setting $\textsl{g} = 8.5 \times 10^{-5}$\,m-s\,$^{-2}$ to match that of Itokawa (cf. Sec.~\ref{sec:algorithm}\ref{sec:slope}), we find $D_{i, \min}\approx 0.015$ m for a rubble  asteroid with properties given in table~\ref{tab:properties}. This result substantiates \cite{miyamoto2006regolith}'s claim that even centimeter-sized impactors may cause global seismic shaking on asteroids like Itokawa.  



The mass loss due to impact ejecta increases with the  impactor diameter. Mass loss occurs  when the velocity of the ejecta exceeds the  escape velocity. To determine the maximum impactor diameter $D_{i,\max}$ that may cause global seismic shaking with  mass loss not amounting to catastrophic disruption, we utilize the commonly invoked parameter $Q^*$ \cite{benz1999catastrophic, melosh1997asteroids}, which is the {\em threshold specific kinetic energy} of the impactor that will precipitate greater than 50\% mass loss. Due to the unavailability of $Q^*$ for a typical asteroid, we take $Q^*$ corresponding to  a target  composed of basalt \citep{benz1999catastrophic}, which has a density similar to that of chondrite. We recognize that for granular targets $Q^*$ may be lower.
To determine the $Q^*$ that corresponds to non-destructive impacts, we utilize the scaling laws of \cite{benz1999catastrophic,melosh1989impact} and limit mass loss to within 1\% by setting $Q^*$ to $Q^*/1000$. Limiting mass loss to within 1\% will avoid introducing complexities due to changes in spin and gravity field; these aspects may be introduced in a more elaborate model in the future. With this, for an asteroid of Itokawa's size, we find that the maximal impactor diameter $D_{i,\max}$  is roughly 1.43m. 

The impactor diameters $D$ are thus constrained to lie  in the range
\begin{equation}
\label{eq:sizerange}
	D_{i,\min} = 0.015\, \text{m} \leqslant D \leqslant 1.43\, \text{m}= D_{i,\max}.
\end{equation}



\subsection{Stochastic collisional history}
To describe the asteroid's collisional history stochastically we  need to define the size distribution of the impactors, as well as the distributions in impact time and in the impact location over the asteroid's surface during its lifetime.

Consider first the size distribution. In Sec. 2\ref{sec:seisprop}, we considered a typical impact to be that between a rubble NEA and an MBO, since most of the NEAs originate in the main belt and spent the majority of their lifetime there before migrating to near-Earth orbits \cite{morbidelli1999origin}. We employ the estimate of \cite{bottke2005linking} to obtain the number of potential impactors over the 10 MYr lifetime of an NEA. Now, a power-law distribution to estimate the total number of MBOs having a diameter greater than $d$ (in km), denoted by $N_{>d}$, is provided by \cite{bottke2005linking} as
\begin{equation}
    \displaystyle N_{>d} = C_d d^{-2.2},
    \label{eq:3.14}
\end{equation}
where the constant $C_d = 2.8 \times 10^6$.  Only a fraction of the MBOs with sizes between $D_{i,\min}$ and $D_{i,\max}$ will collide with a given asteroid. We estimate this fraction as follows. Suppose the total volume of the main belt is $W$ and $N_{>d}$ objects of diameter greater than $d$ exist there. This provides a number density of $N_{>d}/W$. A target asteroid of mean radius $R$ moving with a velocity $U$ will sweep a volume of $\pi R^2 Ut$ in time $t$. Hence, the number of objects $N_h$ that will collide with this asteroid in time $t$ is given by 
\begin{equation}
    \displaystyle N_h = \pi R^2 Ut N_{>d}/W = P_i R^2 N_{>d}t ,
    \label{eq:3.15}
\end{equation}
where $P_i = \pi U/W$ is the {\em intrinsic collisional probability} \cite{jeongahn2017simplified}, which  is $2.18 \times 10^{-18}$  km$^{-2}$ yr$^{-1}$   for collisions between NEOs and MBOs \cite{bottke2005linking}. From  \eqref{eq:3.14} and \eqref{eq:3.15} we then estimate the number of impacts experienced by an asteroid of Itokawa's size and age over its lifetime $t$ of 10 MYr within the size range mentioned in \eqref{eq:sizerange} to be about 150,000.



\begin{figure}
\begin{floatrow}

\capbtabbox{%

	\begin{tabular}{ ccc } 
  \hline
  & Bin 1 & Bin 2 \\ 
  \hline
  Diameter range (m) & 0.015-0.1 & 0.1-1.43 \\ 
  Number of MBOs & 2.80 $\times 10^{17}$ & 1.76 $\times 10^{15}$ \\ 
Number of impacts& 148566 & 938 \\ 
 
 \hline
 \end{tabular}
}{%
  \caption{Number of small (bin 1) and large (bin 2) MBOs, along with the number of expected impacts.}%
\label{tab-2}}
\ffigbox{%
\includegraphics[scale=0.6]{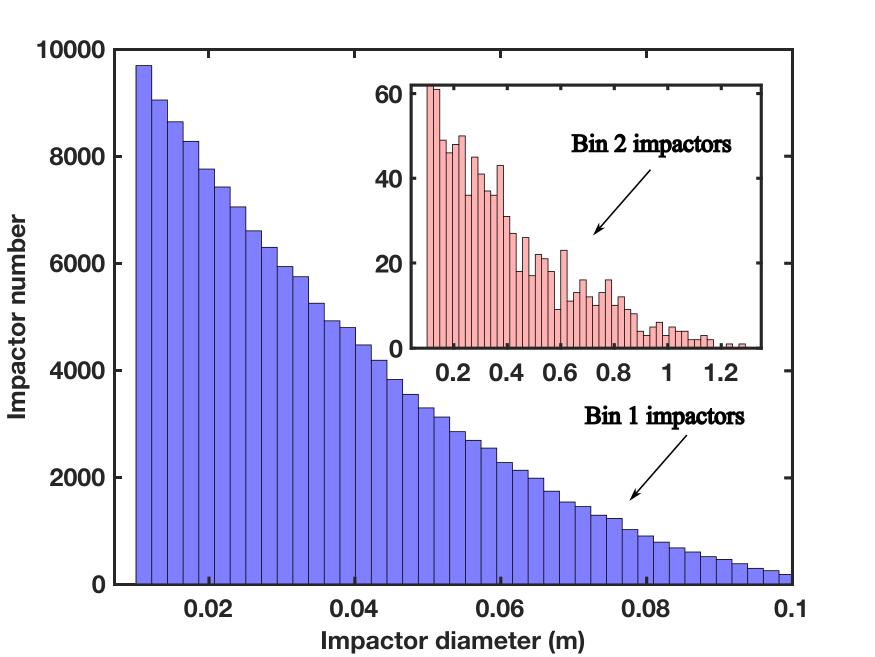} 
}{%
  \caption{Exponential random distribution of impactor diameters in bin 1 and 2. }%
\label{fig:exptdist} 
}
\end{floatrow}
\end{figure}

We now divide the impactor population into two groups or \lq simulation bins\rq $\,$ as shown in table~\ref{tab-2}. {\em Small } impactors in bin 1 have diameters in the range $D_{i,\min}\leqslant D\leqslant D_c$ while {\em large} impactors in bin 2 satisfy $D_c\leqslant D\leqslant D_{i,\max}$,  where the cut-off size $D_c = 0.1$ m is obtained by setting $\Gamma$ at the point of impact to be 100, which corresponds to the modal value of a typical collisional history experienced by our rubble asteroid as seen later in table \ref{meanimpact}.

 {\color{black} Now impactor size will vary within a bin, to describe which we need a size-frequency distribution. Recently, \cite{holsapple2022main} employed a uniform random distribution of impactor sizes within a  bin. Here we utilize an exponential random distribution \cite{stuart2001near}, as we expect the impactor population to reduce with increasing impactor diameter.
 Figure~\ref{fig:exptdist} displays the size distribution we employ for impactors in bin 1 and bin 2.}
 

 
 The surface distribution of impacts  is implemented as follows. We  generate a uniform grid over the ellipsoidal asteroid's surface, as in Fig.~\ref{fig:asteroidX}, and
 then allocate impact sites randomly at the vertices. Similarly, impacts are assumed to have occurred randomly during the asteroid's lifetime, which is equivalent to taking the impactor flux  constant. Finally, combining the size, spatial, and temporal  distributions  we arrive at the necessary stochastic description. 



\begin{figure}
\floatbox[{\capbeside\thisfloatsetup{capbesideposition={right,center},capbesidewidth=6cm}}]{figure}[\FBwidth]
{\caption{Random spatial distribution of impacts on an ellipsoidal asteroid with an area of interest, patch X at a central location. Three points of interest $X_1$, $X_2$, and $X_3$ in this study as seen later in Sec. \ref{sec:results} are at the three ends of the ellipsoidal NEA. }\label{fig:peakgamma}}
{\includegraphics[scale=0.45]{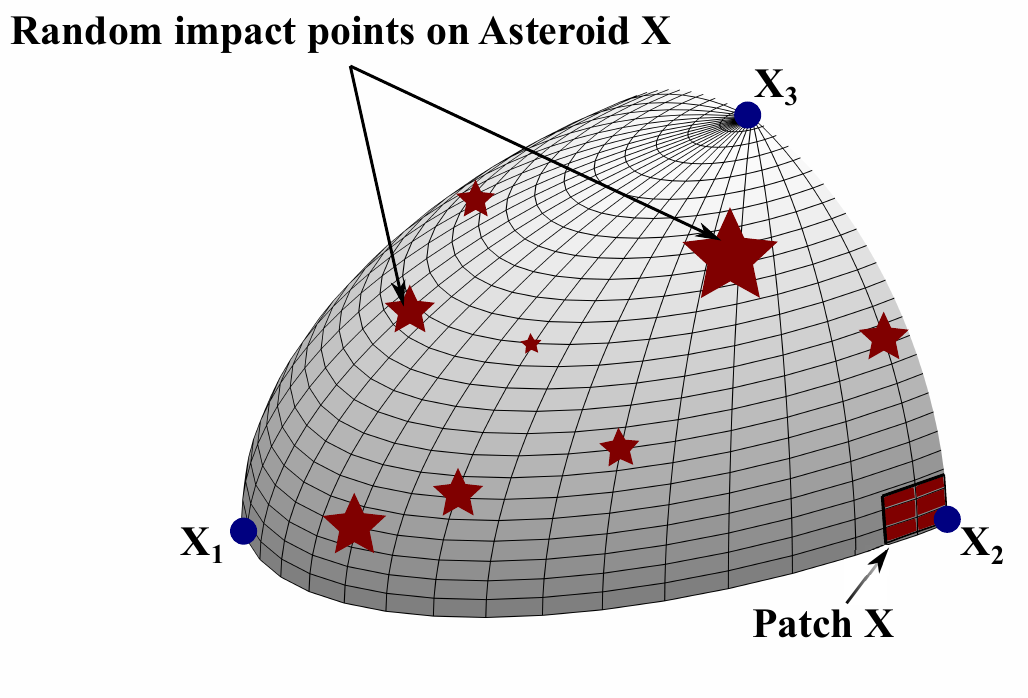}}
\end{figure}

\subsection{Seismic activity at a given point}

\begin{figure}[t!]
\centering
\includegraphics[scale=0.75]{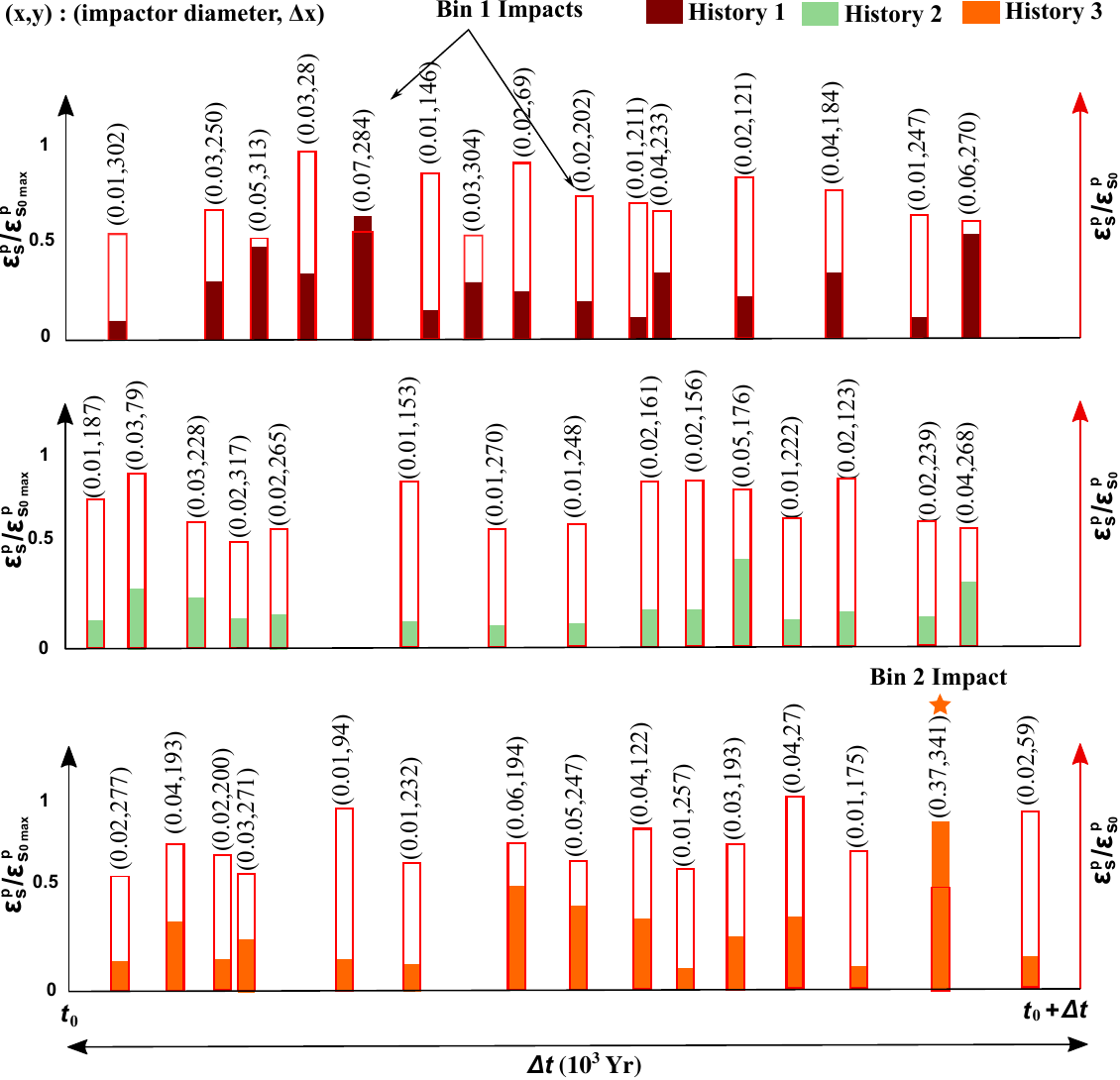}    
\caption{ Scaled peak seismic energy densities  $\varepsilon^p_s/\varepsilon^p_{s0 max}$ (\textit{filled bar}) and $\varepsilon^p_s/\varepsilon^p_{s0}$ (\textit{empty bar}) received at patch $X$ over $10^3$ years for three sample collisional histories.  Each bar represents an impact  at an arbitrarily chosen surface location on the asteroid by an object randomly selected from either bin 1 or bin 2, with a star identifying the latter. The two values in the  parentheses along the abscissa represent, respectively, impactor diameter and distance from point $X$. The horizontal separation between two bars represents the time between two impacts.}
\label{fig:ehistory}
\end{figure}

We are now ready to quantify  the seismic activity  at patch $X$ due to impacts. For this, we will focus on the patch's center, point $X$. Seismic energy  is transferred to point $X$ post-impact, and this  transfer is modeled through \eqref{eq:2.4}. Utilizing this for each impact in a sample collisional history we  compute the seismic energy that is received at $X$ during that particular sequence of impacts. The delivery of peak seismic energy density $\varepsilon^p_s$ at patch $X$ for three different collisional histories over a time span of $10^3$ years is reported in Fig.~\ref{fig:ehistory}. For each impact, we scale $\varepsilon^p_s$ by $\varepsilon^p_{s0max}$ -- the  energy density of the largest impact in the collisional  history containing the impact, as well as  by $\varepsilon^p_{s0}$ --  the energy density at the point of impact. The impactor's size and distance $\Delta x$ from $X$ are indicated in Fig.~\ref{fig:ehistory}.

Because none of the impacts in any of the three histories in Fig.~\ref{fig:ehistory} have $\varepsilon^p_s/\varepsilon^p_{s0max} = 1$, it is clear that the strongest impact in each history occurred outside the $10^3$ year time window considered. We also find that large impacts, i.e. impacts from objects lying in bin 2 in table~\ref{tab-2}, are rare, with only one occurring over $10^3$ years in collisional history 3. Comparing  $\varepsilon^p_s/\varepsilon^p_{s0max}$ across the three histories we find that the distribution of seismic energy density is different, with collision history 2 receiving the least fraction of the maximum $\varepsilon_s$ deposited over the $10^3$ year span. Furthermore, as expected, impactors closer to  $X$ and/or larger in size deliver greater $\varepsilon^p_s$, as they input greater kinetic energy and/or lose less energy during seismic wave transmission to patch $X$. Indeed, by considering $\varepsilon^p_s/\varepsilon^p_{s0}$ in collisional history $1$ in Fig.~\ref{fig:ehistory} we find that the closest impact to point $X$ was only 28m away so that 94.87\% of the seismic energy deposited at the impact location was received at $X$. At the same time, in the same history, the farthest impact from  $X$ was 313m away, so only 50.55 \% of the impact energy was made available at  $X$, with the rest getting distributed and dissipated in the asteroid's interior. Finally, we note that, although Fig.~\ref{fig:ehistory} shows different evolutions of $\varepsilon^p_s$ over $10^3$ years,  the {\em total} seismic energy delivered remains roughly the same for all samples.




We now investigate the non-dimensionalized ground acceleration $\Gamma$ at patch $X$ due to seismic shaking there employing the procedure discussed in Sec.~\ref{sec:richardson}. In Fig.~\ref{fig:peakgamma} we plot $\Gamma$ observed at $X$ due to large impacts  against the distance  $\Delta_X$ of the impact's location from $X$. As expected, $\Gamma$ at point $X$ depends upon both  $\Delta_X$ and the impactor's diameter $D_i$. As with $\varepsilon^p_s$ in Fig.~\ref{fig:ehistory}, impactors that are big in size and/or close to  $X$ produce a high $\Gamma$. In Fig. \ref{fig:peakgamma}, we observe a concentration of impactors lying within 200m-250m  of point $X$, which reflects the central location of patch $X$ on the asteroid; see Fig.~\ref{fig:asteroidX}. Were $X$ located  at the end of the asteroid's long axis, $\Delta_X$ on average would be much higher for a uniform random distribution of impacts on the surface; the consequences of this are  discussed in Sec.~\ref{sec:results}\ref{sec:location}.


\begin{figure}[t!]
\hspace*{-2.5em}
\subfloat[][]{\includegraphics[scale=0.55]{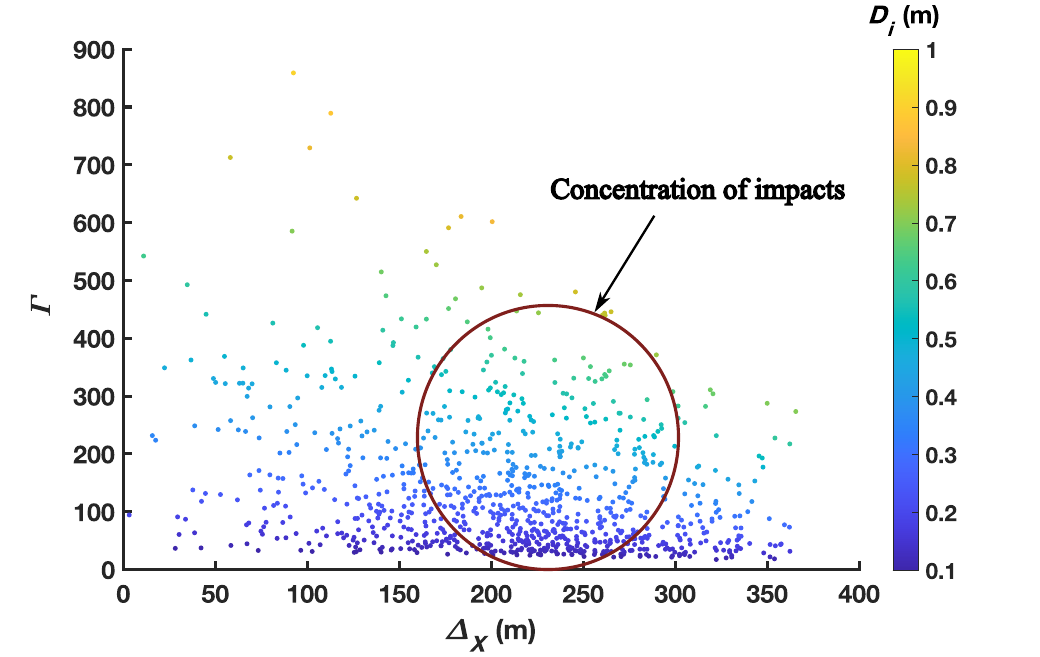} \label{fig:peakgamma}}\hspace*{-2.5em}
\subfloat[][]{\includegraphics[scale=0.56]{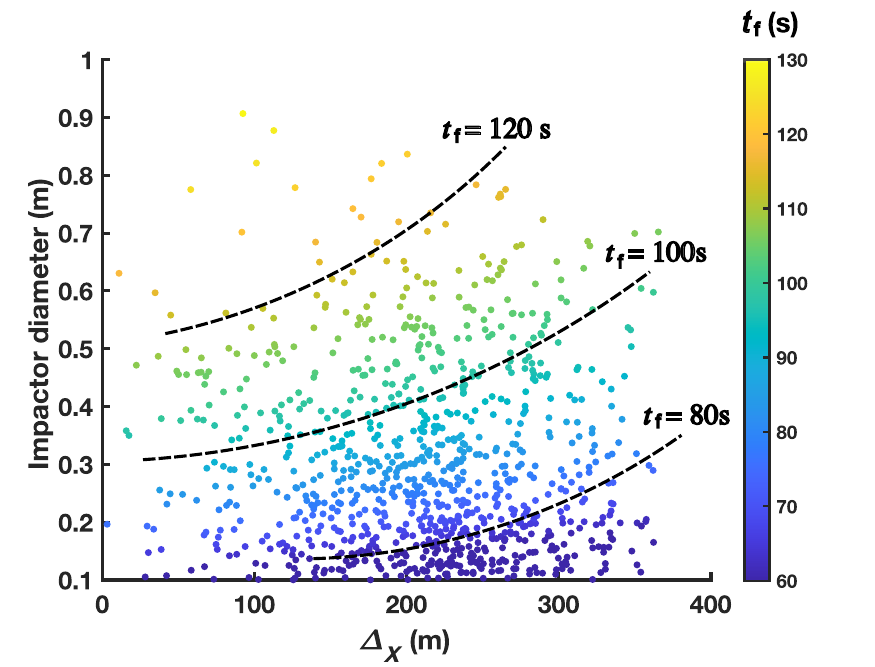}\label{fig:falloff} }
\caption{ (a) Peak ground acceleration $\Gamma$ at patch $X$ as a function of the distance $\Delta_X$ of the impact from $X$ for large impactors, i.e. objects belonging to bin 2. (b) Fall-off time $t_{\text{f}}$ of the seismic energy at point $X$ as a function of $\Delta_X$ for large impacts. Dashed lines represent lines of constant $t_{\text{f}}$. Dots are colored as per the indicated scale. }

\end{figure}

Figure~\ref{fig:signature} demonstrated that, in time, seismic energy received at a location attenuates to zero asymptotically. In reality, due to the presence of rate-independent energy dissipation mechanisms, e.g. dry frictional sliding between grains,  energy will die out in finite time. Such dissipation mechanisms are not present in \eqref{eq:2.4}, which assumes that the  energy loss is due mainly to viscous effects. We nevertheless  incorporate the finite-time dissipation of seismic energy  in our simulations -- cf. Sec.~\ref{sec:algorithm}\ref{sec:dem} --  by introducing a {\em fall-off time} $t_{\text{f}} $, which is defined by  the property that
\begin{equation}
    {E(t_{\text{f}})}= 10^{-4}{E_{\max}}\, ,
    \label{eq:3.16}
\end{equation}
where $E_{\max}$ represents the peak seismic energy experienced at  $X$ due to an impact event.  The fall-off time thus represents the time taken for the seismic energy to decay from its peak value to a small number. In Fig.~\ref{fig:falloff} we display the variation of $t_{\text{f}}$ with both the impactor's diameter and its distance $\Delta_X$ from point $X$. As expected, a large impactor close to $X$ will have the longest $t_{\text{f}}$, while a small impactor far away from $X$ will have the shortest $t_{\text{f}}$.

\section{Methodology}
\label{sec:algorithm}

 It is too computationally expensive to  investigate regolith motion and  segregation over the entire surface of an asteroid  through discrete element (DE) simulations. Thus, we study regolith motion within a region that is large enough to display the segregated state that interests us -- for example,  around the Muses Sea region on Itokawa as shown in Fig. \ref{fig:asteroids} --  but is small enough to be computationally tractable. To this end, we work, as in the foregoing sections, with patch $X$ on the equatorial region of the asteroid in Fig.~\ref{fig:asteroidX}. The patch has dimensions of 20m $\times$ 20m along the $x$ and $y$ directions, respectively, and is taken to be 5m deep in the $z$-direction, consistent with the mean regolith bed depth  estimated by the aforementioned SBCTEM simulations of \cite{richardson2020impact}.


\subsection{Conditions at patch $X$}
\label{sec:slope}\label{sec:grainsize}

To initialize  our simulations we need  the gravitational field and the grain size distribution at patch $X$ in Fig.~\ref{fig:asteroidX}.  Consider first the magnitude of the gravitational acceleration. We set the asteroid's density and rotation period  to  match Itokawa's values, viz. 1.9 g-cm$^{-3}$ and 12.132 h, respectively. Following the methodology of \cite{werner1996exterior} for estimating a polyhedron's gravity field yields $ \textsl{g}_A \approx 8.5 \times 10^{-5}$\,m-s\,$^{-2}$ as the gravitational acceleration  at point $X$. We  take the gravitational field over patch $X$ to be this constant value.  The centrifugal acceleration at the assumed average spin rate of $1.44 \times 10^{-4}\, \text{ rad-s}^{-1}$ is approximately $2.42 \times 10^{-6}$ m-s$^{-2}$. This is about  3\% of the gravitational acceleration at  $X$, so we ignore centrifugal effects henceforth. 

We note that, because of the location of patch $X$ on the asteroid in Fig.~\ref{fig:asteroidX}, gravity acts normally on the surface. This will not be so for less regular asteroid shapes or even at  other latitudes on an ellipsoidal one. At the same time, the local slope -- defined as the angle $\theta$ between the gravity vector and surface normal --  is less than $5^{\circ}$   in the vicinity of the Muses C on Itokawa \cite{scheeres2006actual}. Thus, the vanishing of $\theta$ in our model is representative and not atypical. In fact, when $\theta$ is so much smaller  than the cohesionless angle of repose of the regolith, which for typical geological  materials  lies between 30$^{\circ}-40^{\circ}$  \cite{al2018review}, or even higher in low-gravity environments \cite{elekes2021expression}, a triggering mechanism, such as seismic shaking, will be required to mobilize the grains.

We turn next to the grain size distribution in patch $X$. This is typically non-uniform on asteroids; see, e.g. Fig.~\ref{fig:asteroids}. There are ongoing efforts to model boulder distribution on asteroids, and \cite{tancredi2015distribution} provides one such model for Itokawa obtained by processing  raw images procured by Hayabusa. The prediction of \cite{tancredi2015distribution}'s model for the boulder distribution at Itokawa's eastern face helps set the grain size distribution at patch $X$ in our simulations. We consider three categories of grain sizes in increasing order of their diameters, viz. {\em pebbles} (diameter < 0.1m, labeled \lq S\rq~for {\em small}), {\em cobbles} (diameter $0.1\text{m}-1\text{m}$,  labeled \lq M\rq~for {\em medium}), and {\em boulders} (diameter > 1m,  labeled \lq L\rq~for {\em large}), following the terminology of \cite{miyamoto2007regolith}. In the subsequent sections we refer to  cobbles and boulders  as  {\em M} and {\em L boulders}.

Because of their large numbers, it is  computationally infeasible to simulate all  the smallest grains, also called {\em fines}. The total number of fines  is, therefore, limited to 50000, and we fix their mean diameter to be 0.25 m. This allows us to achieve the desired regolith depth at patch $X$ but keeps computational times reasonable. Table~\ref{tab-gs} displays the grain size distribution that we adopt, with grain sizes distributed uniformly and randomly within 10\% of the given mean diameters. We ignore the presence of very large boulders, i.e. those with diameters greater than 10m; \cite{michikami2021boulder} estimates that there are very few such boulders on Itokawa.



\subsection{Discrete element simulations}
\label{sec:dem}\label{sec:simbox}\label{sec:shakebox}

\begin{figure}
\begin{floatrow}

\capbtabbox{%
\setlength{\tabcolsep}{1pt}
\begin{tabular}{cccc}
\hline
Type & Mean &Number & Volume \%  \\
&diameter (m) & & \\\hline
Pebbles / S fines & 0.25 & 50000 & 95.37 \\
Cobbles / M boulders & 0.75 & 50 & 2.57 \\
Boulders / L boulders & 1.5 & 5 & 2.06 \\\hline
\end{tabular}
}{%
  \caption{Grain distribution at patch $X$  in simulations.}%
\label{tab-gs}}
\ffigbox{%
\includegraphics[scale=0.08]{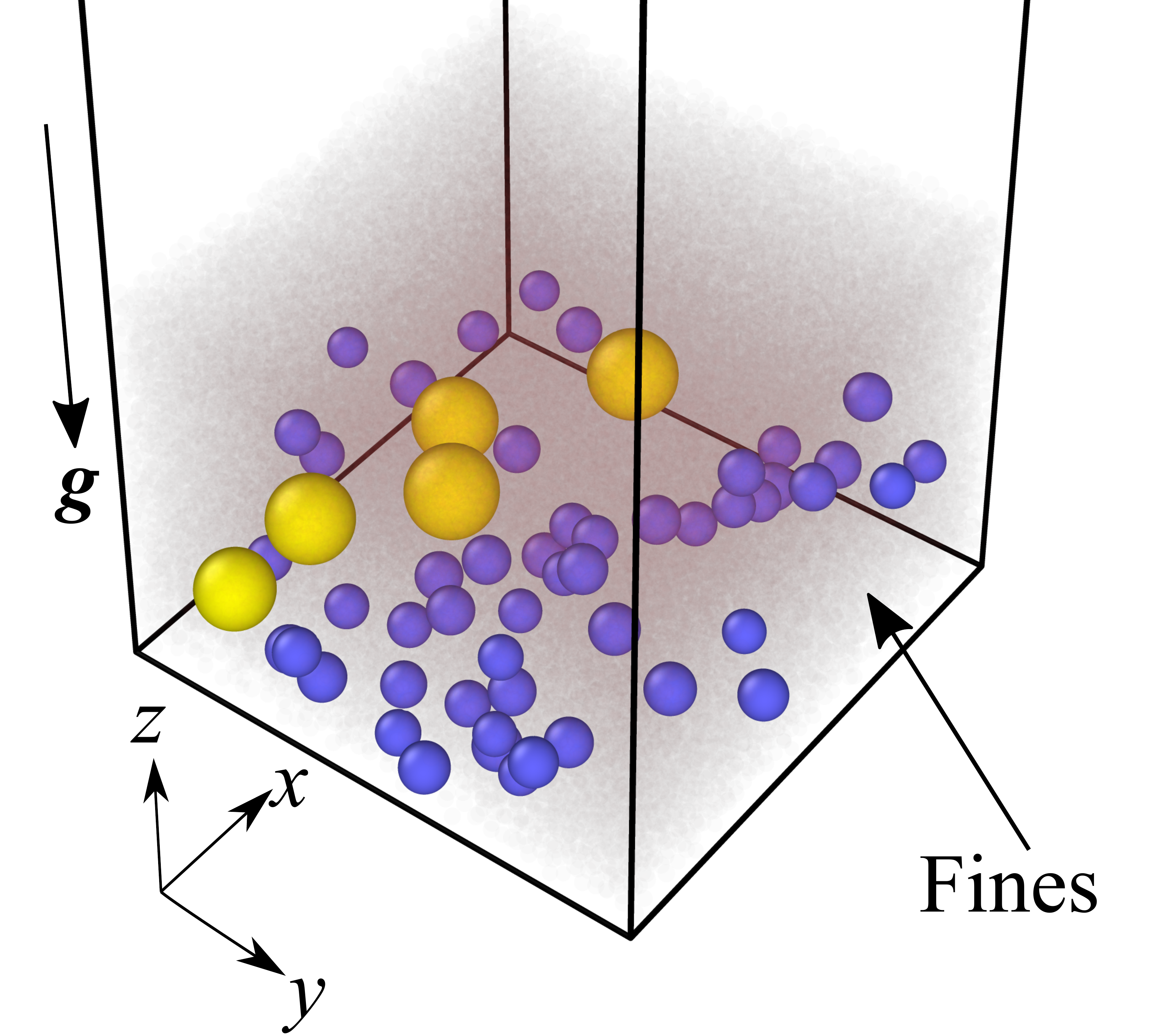} 
}{%
  \caption{The simulation box with different types of grains visualized by OVITO \cite{stukowski2009visualization}. Yellow and blue colors indicate large  (L) and medium boulders (M), respectively. Fines (S) constitute the hazy particulate cloud. The box's orientation relative to patch $X$ in Fig.~\ref{fig:asteroidX} may be ascertained by aligning the axes $x,y$, and $z$.}%
\label{fig:simbox}}
\end{floatrow}
\end{figure}


The motion of grains in patch $X$ is followed through  DE simulations  performed on the open-source package Large-scale Atomic/Molecular Massively Parallel Simulator (LAMMPS) \cite{plimpton1995fast}. In LAMMPS the grains  are taken to be rigid and cohesionless spheres of the same density $\rho =$  3200 kg-m$^{-3}$ corresponding to that of typical LL chondrite, which is believed to be the composition of the regolith on Itokawa \cite{terada2018thermal}. The collisions between grains, or between grains and the walls of the simulation box are inelastic and are described through  interaction models in the normal and tangential directions. These models consist of a non-linear elastic spring placed in parallel to a linear viscous damper, together connected in series with a Coulomb frictional slider; here we suppress the slider in the normal direction. 

The parameters of the elements in the interaction model are tuned to match the overall grain properties as follows.  From Hertzian contact mechanics, the  stiffness of the normal and tangential are obtained as, respectively, $k_n = 5\times 10^4 \text{kg-s}^{-2}$ and $k_t = 2 k_n/7$ \cite{maurel2016numerical}. The viscosity of the damper in the normal direction $\gamma_n$ is tuned by the normal coefficient of restitution $e$, which is set to 0.5,  yielding $\gamma_n = 493 \text{kg-s}^{-1}$ \cite{mishra2003review}, while the viscosity in the tangential direction $\gamma_t=\gamma_n/2$, which is the default value adopted in LAMMPS. We note that $\gamma_n$ and $\gamma_t$ are obtained using a linear spring-dashpot interaction model, but it provides a reasonable first estimate to calibrate our simulations. The angle of friction of the Coulomb slider is taken to be $\Phi = 35^{\circ}$, which corresponds to the typical angle of repose reported for grains on the asteroid Itokawa \cite{fujiwara2006rubble}. Lattice formation is prevented as we allow  grain sizes to vary by  10\% about a mean diameter, as noted above.




As  mentioned at the section's beginning, we will simulate the dynamics of the grains occupying patch $X$ which, as shown in Fig.~\ref{fig:simbox}, is modeled as a cuboidal \lq simulation box\rq~of dimension $20\text{m}\times20\text{m}\times50\text{m}$. Gravity $ \textsl{g}_A \approx 8.5 \times 10^{-5}$\,m-s\,$^{-2}$  acts along the negative $z$-direction. The bottom face in the $z$-direction is a fixed and impenetrable wall and represents either confined bulk grains or the bedrock on top of which the regolith lies.  The top  of the simulation box in the $z$-direction is open and unbounded. Finally, we impose periodic boundary conditions (PBCs) in the lateral $x$ and $y$ directions.  The choice is prompted by both the gravity being  nearly normal to the surface in the vicinity of patch $X$ and our restricting seismic activity to cause shaking along $z$ alone. Thus, patches abutting  patch $X$ may be assumed to respond similarly, at least to the leading order of approximation. 
Employing  PBCs also compensates for the loss of grains that are launched  from patch $X$ to neighboring patches due to impacts, by replacing them with grains that are gained from neighboring patches through a similar process. Assumptions about patch $X$ will have to be revisited at those locations of an asteroid that experience gravity and/or seismic shaking that are not normal to the local topography, and this will be the focus of subsequent work.

To prepare the initial conditions for the simulation, we randomly generate  grains inside the simulation box as per the grain size distribution in table~\ref{tab-gs}. These are then allowed to  settle under gravity, which is temporarily set equal to the Earth's value for faster deposition. The settling time is kept long enough to allow the grains to attain equilibrium. The entire simulation box is then spun about its $y$-axis and then  allowed to resettle to ensure a random initial configuration and to prevent any lingering artifacts from the grain generation process in LAMMPS.

We now investigate the evolution of this mixed collection  of grains at patch $X$ as the asteroid is subjected to collisions over its lifetime. To simulate this the base of the simulation box is made to shake sinusoidally after each impact in a sample collisional history. The walls of the simulation box are smooth and act only to  confine  the grains laterally. As already mentioned, we take seismic shaking to be normal to the topography, which is consistent with our restriction of seismic energy transfer to patch $X$ through only compressional $P$-waves. This is a first step, as actual seismic shaking may also involve motion parallel to the surface, which will be incorporated in future work.



To simulate  seismic shaking at patch $X$ in the aftermath of the $i^{\text{th}}$ impact we need to provide the associated  shaking's duration, frequency and amplitude. The  duration is set by the fall-off time $t_{\text{f},i}$, defined by \eqref{eq:3.16} in terms of the $i^\text{th}$ impact's peak seismic energy $E_{\max, i}$ received at patch $X$.  Figure~\ref{fig:falloff} indicates that the longest fall-off time is about 120s. Then, as discussed at the end of Sec.~\ref{sec:richardson}\ref{sec:seisprop}, we set $f_i=f_0$, the frequency of the first radial mode of vibration estimated in  \eqref{eq:2.14}. Finally, the shaking amplitude is found as follows.  The ground shaking due to the $i^\text{th}$ impact is characterized through the non-dimensional surface acceleration $\Gamma_i$ observed at the patch's center $X$; cf. \eqref{eq:2.4}--\eqref{eq:2.6}.
Employing $\Gamma_i$  as $\Gamma$ in \eqref{eq:2.6} and setting $\omega_i = 2\pi f_i$ we finally compute the  amplitude $A_i$ for the $i^{\text{th}}$ impact.



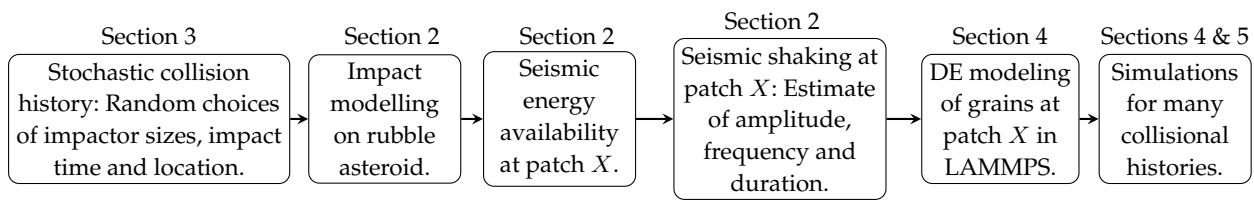
\begin{figure}

\begin{tikzpicture}[node distance=2.0cm]

\node (start) [Level0,align=center] {Stochastic collision \\history: Random choices \\ of impactor sizes, impact \\time and location.};


\node (in2) [Level0, right of=start,xshift=1.1cm,align=center] {Impact \\modelling \\ on rubble \\ asteroid. };

\node (in3) [Level0, right of=in2,xshift=0.3cm,align=center] {Seismic \\energy \\availability \\at patch $X$.};

 \node (in4) [Level0, right of=in3,xshift=0.9cm,align=center] {Seismic shaking at\\ patch $X$: Estimate\\  of amplitude,\\ frequency and\\ duration. };

 \node (in5) [Level0, right of=in4,xshift=0.9cm,align=center] {DE modeling\\ of grains at\\ patch $X$ in\\ LAMMPS.};

 \node (in6) [Level0, right of=in5,xshift=0.3cm,align=center] {Simulations \\ for many \\ collisional \\ histories.};

 \draw [arrow] (start) -- (in2);
 \draw [arrow] (in2) -- (in3);
 \draw [arrow] (in3) -- (in4);
 \draw [arrow] (in4) -- (in5);
 \draw [arrow] (in5) -- (in6);

 \draw (0,1.1) node [] {Section 3}; 
  \draw (3.2,1.1) node [] {Section 2}; 
  \draw (5.5,1.1) node [] {Section 2}; 
  \draw (8.2,1.3) node [] {Section 2}; 
  \draw (11.2,1.1) node [] {Section 4};
   \draw (13.5,1.1) node [] {Sections 4 \& 5};


\end{tikzpicture}
\caption{Modeling flowchart for segregation from impact-induced seismic shaking at patch $X$ on a rubble  asteroid. } 
\label{fig:flowchart}
\end{figure}

The effect of any impact on the regolith at patch $X$ may now be found by simulating in LAMMPS the shaking of the granular mixture kept in the simulation box --   described above and shown in Fig.~\ref{fig:simbox} -- at the impact's associated amplitude and frequency, and over the corresponding fall-off time.  Repeating this process for {\em every} impact  in {\em any} sample collisional history of the asteroid then allows us to investigate impact-driven segregation at patch $X$ over the asteroid's lifetime.  Figure~\ref{fig:flowchart} summarizes the entire modeling process in a flowchart. 

\begin{figure}[t!]
\centering
\includegraphics[scale=0.75]{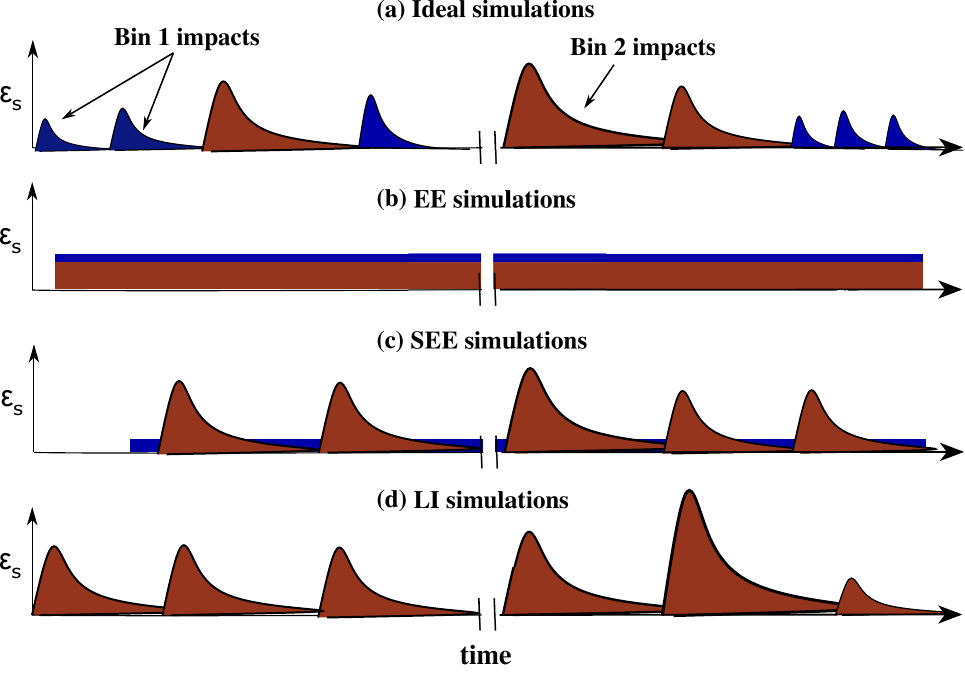}
\caption{(a) Schematic showing the sequence of seismic energy curves representing energy transfer to the center of patch $X$ in a  collision history, with blue and red corresponding to small (bin 1) and large (bin 2) impacts, respectively. (b) \textit{EE simulations}: energy transfer from both large and small impacts are smeared out. (c) \textit{SEE simulations}: energy transfer from large impacts is modeled individually, while that from small impacts is smeared out and superimposed. (d) \textit{LI simulations}: Same as SEE, but with small impacts ignored. See the main text for more details. The figure is not to scale.}
\label{fig:eqsim}
\end{figure}

As mentioned, post-impact shaking at patch $X$ may last at most 120s. There is, therefore, a long period of time between successive impacts during which the grains in patch $X$ remain static. It is possible that grains during such periods may experience creeping flow \cite{ballouz2019surface} or be subjected to perturbations due to internal or external events, like degassing, quakes, or radiation torques. We do not take into account these possibilities here. This permits us to collapse the impact-driven grain evolution at  patch $X$ into a concatenated sequence of shaking events, whose chronology is set by the sample collisional history under consideration. Thus, shaking at patch $X$ due to the $(i+1)^{\text{th}}$ impact is initiated as soon as the simulation associated with the $i^{\text{th}}$ impact terminates. Figure~\ref{fig:eqsim}(a) displays a schematic of how a collisional history would ideally be simulated; we discuss the figure further below. 


\subsection{Energy equivalent simulations}
Because of the large number of impacts ($\sim150,000$), simulating even one collisional history requires excessive computational time. 
We overcome this in three ways. In the first, we perform one {\em energy-equivalent} (EE) simulation for each collisional history. The equivalence is established on the basis of the {\em total} seismic energy received at patch $X$ from all impacts in a sample collisional history of the asteroid. The  seismic energy delivered to the center of patch $X$ by an impact is the area under the associated seismic energy curve, such as the one in Fig.~\ref{fig:signature}. The idea of EE simulations is further  explained in Fig.~\ref{fig:eqsim}. Figure~\ref{fig:eqsim}(a) represents an ideal simulation in which each collision -- small or large -- is resolved. Collisions with bin 1 (small) and 2 (large) impactors are colored blue and red, respectively. Figure~\ref{fig:eqsim}(b) is an EE simulation in which energy is delivered to patch $X$ through an equivalent sinusoidal shaking with a {\em fixed} amplitude $A_{E}$ and a {\em constant} frequency $f_{E} = f = f_0$ for a given length of time $\Delta t_{E}$. The energy transfer is thus represented as a straight line in Fig.~\ref{fig:eqsim}(b), the area under which equals the total seismic energy received in Fig.~\ref{fig:eqsim}(a). Sections of the area in Fig.~\ref{fig:eqsim}(b) are colored differently to identify energy received from small and large simulations separately; the areas of which equal the total areas under the corresponding curves in Fig.~\ref{fig:eqsim}(a).  

 
To implement an EE simulation we require the duration $\Delta t_E$ and  amplitude $A_E$ of the  shaking, which we now estimate. The energy transfer rate per unit mass -- i.e. the specific power -- from the equivalent sinusoidal motion is
\begin{subequations}
\label{eq:4.22}
\begin{flalign}
    && \zeta_{E} = 4\pi^2A_{E}^2f_{E}^3, && \label{eq:4.22a}\\
&\text{so that we must have}  &\zeta_{E} \Delta t_{E} = \sum_{i = 1}^N \int_0^{t_{f,i}} \varepsilon_{s,i}\text{d}t,&\qquad&
    \label{eq:4.22b}
\end{flalign}
\end{subequations}
where $\varepsilon_{s,i}$ represents the seismic energy density at patch $X$ following the $i^{\text{th}}$ impact with fall-off time $t_{f,i}$.  Now, recall that the frequency $f_E$ is fixed at that of the body's first mode of vibration, i.e. $f_0$. We may then obtain $A_E$ for any collisional history in terms of an associated equivalent scaled surface acceleration $\Gamma_{\text{eq}}$ by utilizing definition \eqref{eq:2.6}; we discuss below how $\Gamma_{\text{eq}}$ is  estimated. Finally, we  utilize \eqref{eq:4.22a} to find the specific power $\zeta_E$, and then \eqref{eq:4.22b} to compute $\Delta t_E$ from a given collisional history. All past work on  segregation on asteroids due to shaking utilize, in effect,  EE simulations \cite{tancredi2012granular,matsumura2014brazil,maurel2016numerical}, although no effort is made to relate the shaking parameters to a geophysical process such as impacting, as done here. 

We now return to estimating the equivalent  surface acceleration $\Gamma_{\text{eq}}$ for a collisional history, which was required in the preceding paragraph. Table~\ref{meanimpact} lists some choices of  $\Gamma_{\text{eq}}$  and the corresponding $\Delta t_E$ and $A_E$. The first choice in table~\ref{meanimpact} is the  average of the peaks of the scaled surface accelerations $\Gamma_i$  experienced at patch $X$  due to all the impacts in a  collisional history, i.e. $ \Gamma_1 = \sum_{i=1}^N\Gamma_i/N$, where $N$ is the total number of impacts. The second choice for $\Gamma_{\text{eq}}$ in table~\ref{meanimpact} is $\Gamma_2=\left(\max_{i}\Gamma_i+\min_{i}\Gamma_i\right)/{2},$
where $i$ ranges  over all impacts in the  collisional history. The third choice $\Gamma_3$ in table~\ref{meanimpact} is the average over ten collisional histories of the modal values of the peak accelerations $\Gamma_i$ of each his. For reasons discussed in Sec.~\ref{sec:results}\ref{sec:smalllarge} we we will set $\Gamma_{\text{eq}} = \Gamma_3$, for which the amplitude $A_E$ is about twice the smallest grain diameter.

\begin{table}[!t]
\centering
\begin{tabular}{cccc}
\hline

 $\Gamma_{\text{eq}}$ &  $\Delta t_E$ (years) & $f_E = f_0$ (Hz) &  $A_E/d$\\
\hline
$\Gamma_1$ = 4.20 & $1.70 \times 10^4$& 0.0197&0.09\\
$\Gamma_2$ = 330.02 & 2.77 &0.0197&7.32  \\

$\Gamma_3$ = 101.35 & 26.15 & 0.0197 & 2.25\\
\hline
\end{tabular}
\caption{Various choices of $\Gamma_{\text{eq}}$ for EE simulations and the corresponding shaking time $\Delta t_E$, frequency $f_E$ and amplitude $A_E$ in terms of  small grain diameter.  }
\label{meanimpact}       
\end{table}



Unfortunately, the influence of the stochasticity of the collisional history is lost when we perform an EE simulation. Furthermore, for any sample collisional history, only about 15$\%$   of the total seismic energy available is contributed by small impacts with objects from bin 1, despite their much greater numbers; cf. table~\ref{tab-2}. Figures~\ref{fig:peakgamma} and \ref{fig:falloff} show that both the peak value of the surface acceleration and the time for which the shaking is sustained are higher for large impacts. This points to the need to incorporate the role of large impacts more carefully in simulations. To correct to an extent for the loss of stochasticity in EE simulations, and to better model large impacts, we  pursue two other types of simulations, viz. {\em semi-energy-equivalent} (SEE) and {\em large-impact} (LI) simulations. 

In SEE simulations we approximate the energy received by the small impacts, i.e. those  from bin 1 impactors  in table~\ref{tab-2}, in terms of an energy-equivalent sinusoidal shaking. However,  each large impact from objects in bin 2 in table~\ref{tab-2} is incorporated individually. The energy-equivalent simulations for the small impacts run in the background as shown in Fig. \ref{fig:eqsim}(c): a gentle continuous tapping providing energy equal to all small impacts in a collisional history -- indicated by the blue rectangular box -- is superimposed on seismic shaking from each large impact in that  history. Although computationally more expensive, SEE simulations will help us understand the role and importance of the randomness of the impacts. Further, we will better appreciate the significance of fewer but larger impacts to segregation on an asteroid's surface relative to the  greater number of small impacts. 

Finally, in LI simulations we ignore  small impacts altogether but consider each large impact in any collisional history individually.


\section{Results and discussion}
\label{sec:results}
We now present the results of our discrete element (DE) simulations that investigate segregation at patch $X$ due to impact-driven seismic shaking over the course of the past 10 Myr of a rubble NEO like Itokawa. Specifically, we consider the triaxial ellipsoid in Fig. \ref{fig:asteroidX} with dimensions 535m $\times$ 294m $\times$ 209m with patch $X$ located as shown. For this, we bring together  various aspects of the process described in the preceding sections and displayed in the flowchart in Fig.~\ref{fig:flowchart}, viz. a stochastic collisional history, impact modeling on the asteroid, seismic energy transmission to  patch $X$, grain distribution at patch $X$ and, finally, simulation strategy and parameters.  The degree of segregation is quantified by counting the number of medium (M) and large (L) boulders  that rise to the top  relative to their total numbers. Complete segregation is said to have occurred when all M and L boulders come to the  top of the mixture.




 \subsection{Effect of small and large impacts}
 \label{sec:smalllarge}
\begin{figure}[b!]
\begin{center}
\subfloat[][]{\includegraphics[scale=0.8]{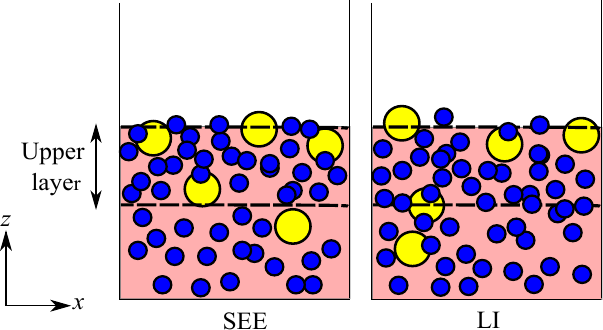} \label{fig:partA12}}
\subfloat[][]{\includegraphics[scale=0.8]{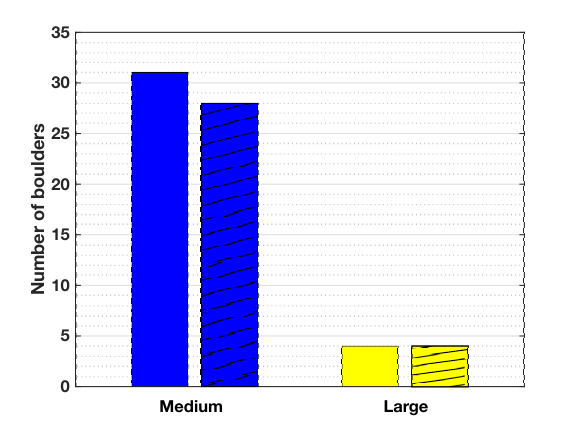}\label{fig:partA3} }
\caption{The final  configuration at the end of  impact-driven shaking  for a sample collision history. Side view of the simulation box for (a) SEE simulations and LI simulations. Only medium (blue) and large (yellow)  boulders are shown. (b) The distribution of medium and large boulders present in the upper layer in SEE (unhatched) and LI (hatched) simulations.}
\end{center}
\end{figure}

Consider first the energy-equivalent (EE) simulations.  The choice of the equivalent surface acceleration $\Gamma_{\text{eq}}$ in EE simulations plays a crucial role in determining whether or not segregation occurs. For example, when we set $\Gamma_{\text{eq}} = \Gamma_1$ given in table \ref{meanimpact}, the shaking amplitude comes out to be much lower than the smallest grain diameter $d$, as there are far more small impacts than large ones -- cf. table~2 and Fig.~\ref{fig:peakgamma}. On the other hand, when we set $\Gamma_{\text{eq}} = \Gamma_2$, the shaking is too violent and grains fly away from  the simulations box. 
However, for $\Gamma_{\text{eq}}=\Gamma_3$, grains are  displaced by an amount of the order of or greater than $d$, and we observe  that big grains rise up imitating the Brazil-nut effect (BNE), and similar to images in Fig.~\ref{fig:partA12}.  This agrees with the findings of \cite{maurel2016numerical} that BNE only occurs when the amplitude of vibration is at least of the order $d$ and confirms the sensitivity of segregation to the amplitude of seismic shaking. 


 We next perform semi-energy-equivalent (SEE) and large-impact (LI) simulations and the final results of these for a sample collisional history are shown in Fig.~\ref{fig:partA12}. As mentioned in table \ref{tab-gs}, initially there are 50 medium and 5 large boulders in the mixture. At the end of the seismic shaking process, the  bar graph in Fig.~\ref{fig:partA3} reports that both M and L boulders predominantly populate the upper layers (UL) at patch $X$; the UL is taken to span the first 3m from the surface, i.e. about twice the large particle diameter. The almost similar outcomes of SEE and LI simulations confirm that large impacts are the primary drivers of segregation. This is expected because, as discussed above, large impacts deliver most of the seismic energy to patch $X$ and result in  higher amplitudes of ground vibration that last longer as the fall-off time  $t_{\text{f}}$ is greater. Furthermore, because the difference in the outputs of SEE and LI simulations is minor, it suggests that the effect of small impacts  on the segregation process is minimal and may be ignored, so we may limit ourselves to LI simulations. Nevertheless, we will here continue with SEE simulations to model impact-induced seismic shaking and grain segregation processes.


 \subsection{Effect of location of patch $X$}
 \label{sec:location}


\begin{figure}[t!]
\begin{floatrow}

\capbtabbox{%
\setlength{\tabcolsep}{2pt}
\begin{tabular}{cccc}

\hline
   && $\int_i \epsilon_s\text{d}t\, (\text{J-s-kg}^{-1}) $ & \\ \hline &$X_1$&$X_2$&$X_3$\\ 
\hline
Bin 1  & 5.31 $\times$ $10^{5}$ & 6.13 $\times$ $10^{5}$ & 6.35 $\times$ $10^{5}$ \\
Bin 2  & 3.38 $\times$ $10^{6}$ & 3.87 $\times$ $10^{6}$ & 3.98 $\times$ $10^{6}$\\
\hline
\end{tabular}
}{%
  \caption{The seismic energy received at three different locations of patch $X$ for  the same collisional history.}%
\label{tab:enrec}}
\ffigbox{%
\includegraphics[scale=0.55]{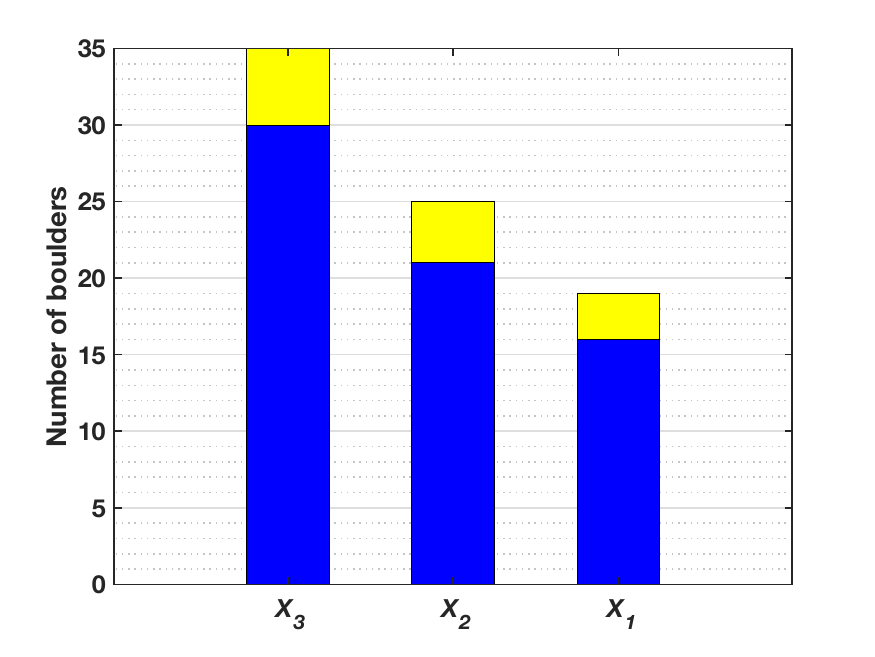} 
}{%
  \caption{The number of M (blue) and L (red) boulders in the UL in the three patches defined in \ref{tab:enrec}. The asteroid is subjected to the same collisional history in all cases.  }%
\label{fig:partB}}
\end{floatrow}
\end{figure}


 The location on the asteroid  of patch $X$ will affect the segregation there. This is because, over the asteroid's lifetime, the spatial distribution of impacts  relative to patch $X$  depends upon patch $X$'s location, and this, in turn, will modify the seismic energy received at $X$. 
To investigate this we compute the total seismic energy received at patch $X$ due to the {\em same} collisional history for three different locations of patch $X$ on the ellipsoid asteroid in Fig. \ref{fig:asteroidX}. These locations correspond to the extremes of the three principal axes, so that patch $X$'s  center coincides with $X_1: (535/2, 0, 0), X_2: (0, 294/2, 0)$ and $X_3: (0, 0, 209/2)$.

As shown in table~\ref{tab:enrec}, for a given collisional history, the seismic energy received is most when the patch is more centrally located, i.e. at the end of the shortest principal axis ($X_3$), while the patch at the end of the longest principal axis ($X_1$) is delivered the least seismic energy. This is expected because seismic waves have to, on average, travel a smaller distance to reach $X_3$ and, consequently, dissipate less. We may quantify the travel extent as the  weighted average distance in a collisional history of the large impactors from a patch, which we define by $d_I = \sum \Delta x_i D_i/\sum D_i$, where $D_i$ is the impactor diameter which strikes at a distance $\Delta x_i$ from the patch. We find that for $X_1, d_I  = 298$m, at $X_2, d_I = 218$m and for $X_3, d_i =190$m. Correspondingly, the seismic energy received at central locations on the rubble asteroid  due to impacts is higher than in peripheral areas. Consequently, in SEE simulations conducted for the  same large impact history, and employing the {\em same} equivalent scaled surface acceleration $\Gamma_3$  for the superimposed small impacts,  we observe that greater numbers of M and L boulders rise to the top when the patch is at the end of the shorter principal axis, i.e. at $X_3$, as shown in Fig.~\ref{fig:partB}.



The above result has important implications for elongated asteroids, such as Itokawa, where we would then expect higher segregation at central locations than at the far ends. However, for more spherical asteroids like Bennu and Ryugu, we predict similar texturing everywhere over the surface.

\subsection{Effect of initial conditions and collisional history} 

\begin{figure}[t]
\begin{center}
\subfloat[][]{\includegraphics[scale=0.8]{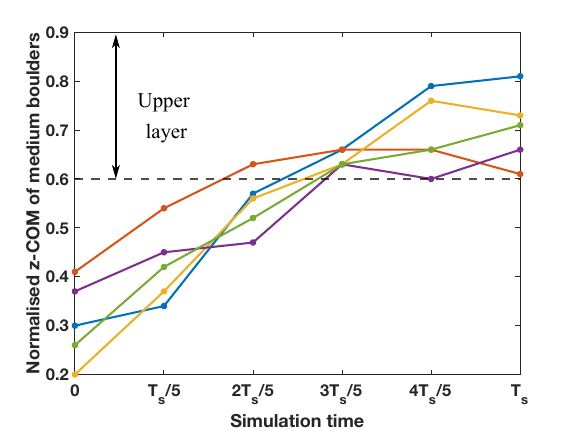} \label{fig:difficA}}
\subfloat[][]{\includegraphics[scale=0.8]{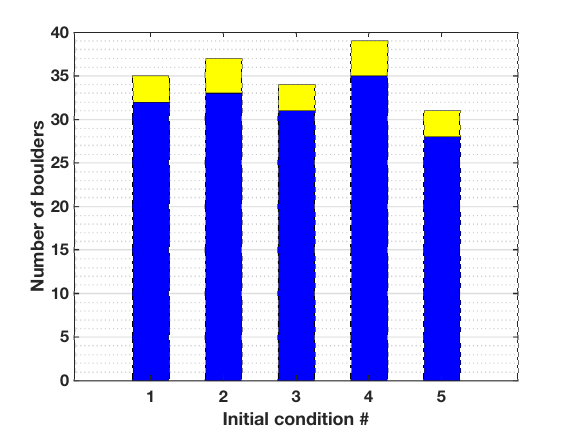}\label{fig:difficB} }
\caption{Outcomes of five simulations with different initial grain configurations, but the same collisional history. (a) Temporal evolution of the vertical location of the center-of-mass (z-COM) of M boulders normalized by regolith depth, with $T_s$ being the total simulation time. (b) The final number of M (blue) and L (yellow) boulders in the UL. }
\end{center}
\end{figure}

\begin{figure}[t!]
\begin{center}
\subfloat[][]{\includegraphics[scale=0.8]{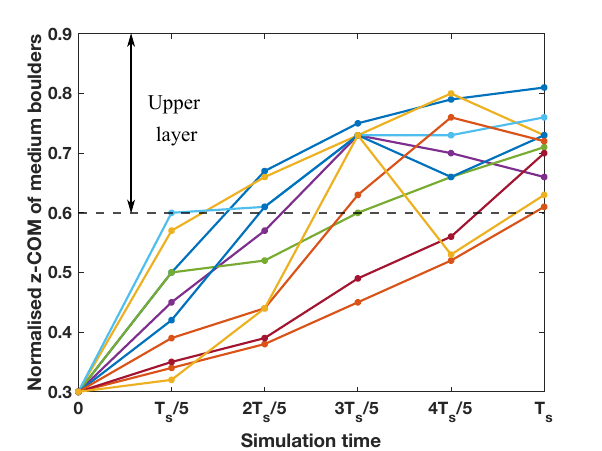} \label{fig:diffchA}}
\subfloat[][]{\includegraphics[scale=0.8]{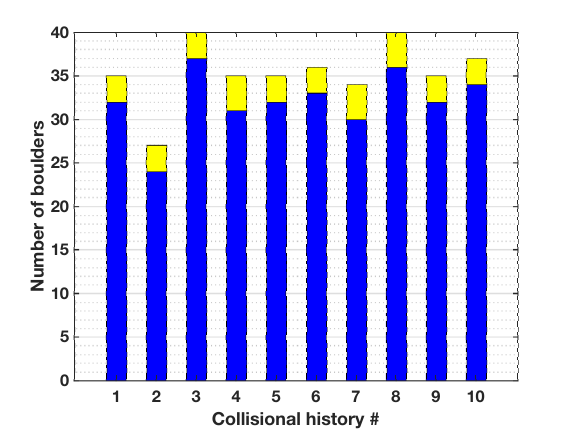}\label{fig:diffchB} }
\caption{Outcomes of ten simulations with different collisional histories, but same initial grain configurations. (a) Temporal evolution of  the vertical location of the center-of-mass (z-COM) of M boulders normalized by regolith depth, with $T_s$ being the total simulation time. (b) The final number of M (blue) and L (yellow) boulders in the UL. }
\end{center}
\end{figure}

We first investigate whether the final outcome is affected by the initial state of the grains in patch $X$. For this, we carried out five SEE simulations for the {\em same} collisional history, but with different initial mixture configurations.  We find that in all these  simulations, although the initial normalized vertical location of the center-of-mass (z-COM) for the M boulders differs across configurations, as shown in Fig.~\ref{fig:difficA}, the M and L  boulders rise to the UL when seismically shaken; cf. Fig.~\ref{fig:difficB}. This confirms the repeatability of segregation and its independence from variability in initial conditions. 

We next study how segregation in patch $X$ is affected by stochasticity in the large impacts through SEE simulations. Figure~\ref{fig:diffchA} reports the outcome from ten different collision histories for the {\em same} initial state of grains. From these simulations, we observe that the temporal evolution of the z-COM of the M boulders in Fig.~\ref{fig:diffchA} depends upon the history of large impacts. However, as  reported in Fig.~\ref{fig:diffchB}, the final segregated state does not change much: the UL is dominated by M and L boulders, although a few histories are segregated to better (histories 3 and 8) or lesser (history 2) extent. This result shows that grain segregation depends primarily upon the total seismic energy delivered by the large impacts, and is largely independent of their collisional history. Having said that, because the evolution of the boulders does vary with the details of the collisional history, it is possible that in the presence of other processes sensitive to surface features, such as YORP,  the outcome could depend more significantly on  stochasticity in the collisional history. Furthermore, robustness to the choice of distribution functions that enter into the stochastic description has yet to be confirmed.

 \subsection{Effect of  asteroid properties}


 \begin{figure}[t!]
\centering
\includegraphics[scale=0.8]{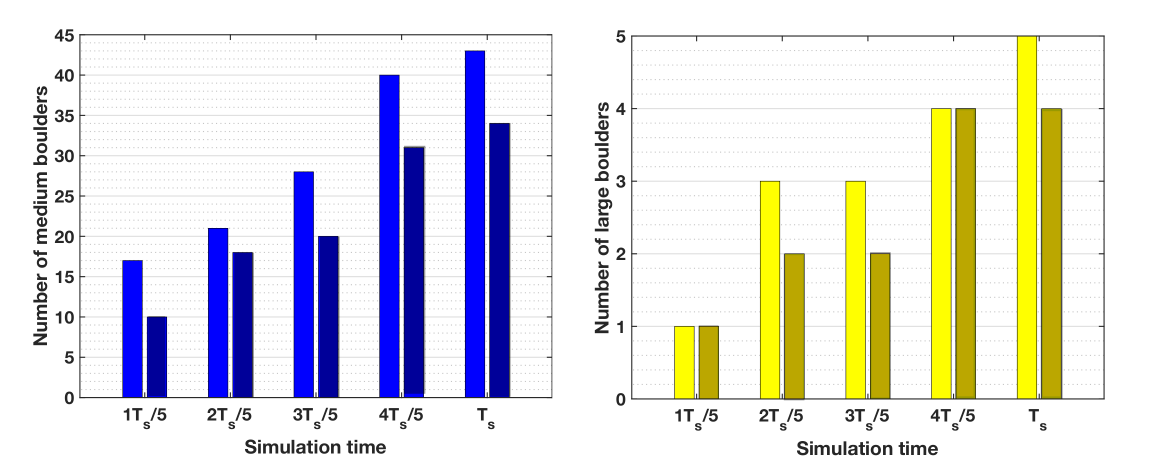}
\caption{The number of M (left, blue) and L (right, yellow)  boulders in the UL of patch $X$ at different simulation time-steps $T_s$ for $K_s = 0.002$ km$^2$ s$^{-1}$ (dark shade)  and $K_s = 0.5$ km$^2$ s$^{-1}$ (light shade). }
\label{fig:partD}
\end{figure}


We now perform SEE simulations to understand the effect of bulk properties of an asteroid on seismic energy propagation and, consequently, on segregation at patch $X$. For this, we compare the present results with that of another asteroid, which is exactly the same in all respects, except that it has a  seismic diffusivity $K_s = 0.5$ km$^2$ s$^{-1}$, equal to that of Eros \cite{richardson2020impact}. This higher value of $K_s$  corresponds to an interior that imitates a fractured monolith, rather than the  much more dissipative rubble pile that we have focussed upon so far. 
 When varying $K_s$ we also need to modify $v_p$ accordingly. The relationship between $K_s$ and $v_p$ for a medium that is not completely monolithic is presently not known so we rely on the analysis in Sec.~2\ref{sec: seismicenergypropagation} to estimate  $v_p$. In Sec.~2\ref{sec: seismicenergypropagation} we found that $v_p$ is about 50\% of the value obtained from \eqref{eq:2.8}; this then yields $v_p = 2.5$ km/s for $K_s = 0.5$ km$^2$ s$^{-1}$. This should be compared with $v_p = 8$ m-s$^{-1}$ computed for $K_s = 0.0002$ km$^2$ s$^{-1}$ in Sec.~2\ref{sec: seismicenergypropagation}. When $K_s$ is greater a larger amount of seismic energy is received at patch $X$, and we expect that this significantly reduces the time taken for  segregation, as well as improves its quality.  This indeed is what we observe in Fig.~\ref{fig:partD} that compares the number of  boulders in the UL. Nearly all the M  and L boulders occupy the UL when the interior is the less dissipative fractured monolith, which reduces their population when the interior is rubble.


 
\begin{figure}[t!]
\begin{center}

\subfloat[][]{\includegraphics[scale=1]{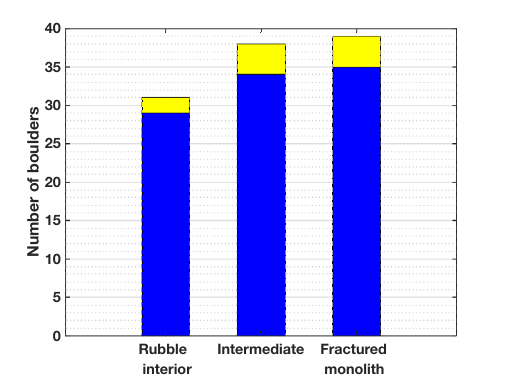} \label{fig:grancompetentA}}
\hspace*{-1.9em}
\subfloat[][]{\includegraphics[scale=0.6]{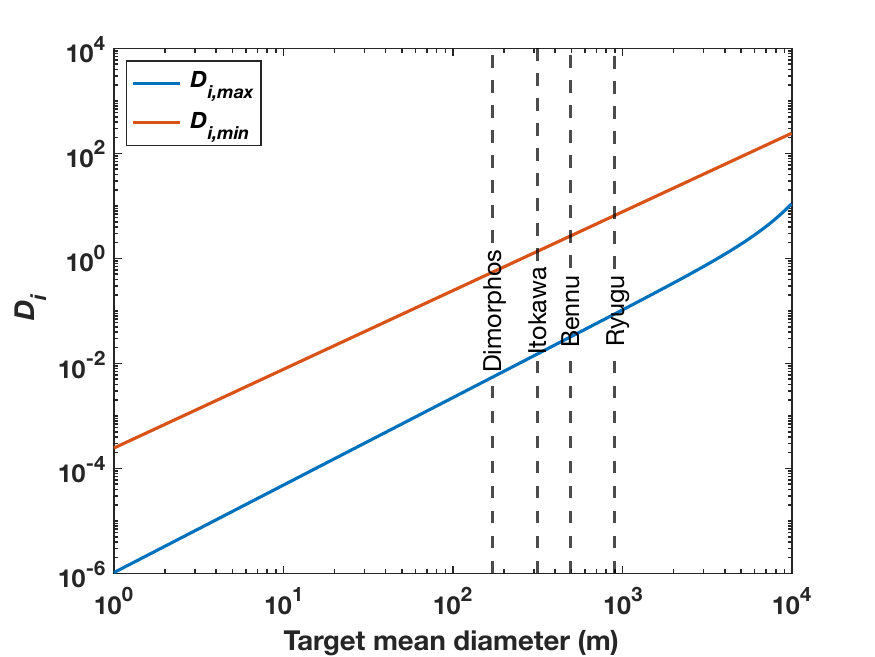}\label{fig:grancompetentB} }
\caption{ (a) The number of medium (M, blue) and large (L, yellow) boulders in the UL for asteroids with interiors characterized by different $P$-wave velocities.  (b) Variation of $D_{i,\min}$ and $D_{i,\max}$ for different target diameters. }
\end{center}
\end{figure}

 To further investigate how the asteroid's interior affects segregation, we performed SEE simulations with   asteroid  interiors that range from  fractured monolith to  granular. We characterize the latter two by  different P-wave velocities: $v_p = 1 \text{km-s}^{-1}$ for a fractured monolith and $v_p = 10 \text{m-s}^{-1}$ for a rubble interior. Additionally, we investigate an asteroid with an interior whose  $v_p = 100 \text{m-s}^{-1}$, an intermediate value.  The associated seismic frequencies  estimated in Sec.~\ref{sec:richardson} will also be different for the three cases.  Other seismic parameters, the asteroid's size, its collisional history, and initial conditions are kept the same. For all  three types of asteroids, Fig.~\ref{fig:grancompetentA} reports that most of the M and L boulders lie in the UL at the end of the collision-driven evolution. Further, the degree of segregation in a fractured monolith asteroid is more than in a rubble-pile asteroid. However, much difference is not observed between asteroids with a fractured monolith and intermediate interiors. This is related to the fact that our asteroid was taken similar in size to  Itokawa. Consequently, seismic energy from even the furthest impact was able to reach patch $X$, notwithstanding the difference between a fractured monolithic or intermediate interior. This may not happen for a larger asteroid, in which the difference between these two interiors may be starker. This aspect leads us to consider briefly the effect of the asteroid's size on segregation.

 Increasing the asteroid's size would require us to recalculate the number of potential impactors, their sizes, and the mean collisional lifetime, and then repeat the entire methodology outlined in this work. For example, following Sec.~\ref{sec:holsapple}\ref{sec:limits} we may estimate the minimum ($D_{i,\min}$) and maximum ($D_{i,\max}$) permissible impactor diameters as a function of the asteroid's size. As Fig.~\ref{fig:grancompetentB} reports both $D_{i,\min}$ and $D_{i,\max}$ rise with the size of the asteroid. This, in turn, will affect the number of impacts that the asteroid will experience over its lifetime: while a larger asteroid will attract more impacts, the number of consequential impacts -- i.e. those that are large enough to initiate seismic shaking --  will reduce. Table \ref{tab:noofimpacts} estimates these numbers for some asteroids. We find that there is a large variability in the number of consequential impacts. This, and the discussion in subsection \ref{sec:location} above, suggests that regolith motion driven by impact-induced seismic activity may need to be investigated with greater attention to the geometric details of an asteroid. These issues will be investigated in a future work utilizing the broad framework presented here. 




\begin{table}

\begin{tabular}{ cccc } 
  \hline
 Target &Mean  & Mean collisional & Number of impacts such that  \\ 
 Body&diameter (km)& lifetime (Yrs)& $D_{i,\min} \le D \le D_{i,\max}$\\
  \hline
   Itokawa & 0.313 & $10^7$ \cite{bottke2005linking} & 149504\\
     Ryugu & 0.900 & $10^7$ \cite{bottke2005linking} & 60893\\

  Gaspra  & 12.2 & $10^8-10^9$ \cite{farinella1992collisional}& 85-845 \\ 
Mathilde & 52 & $4 \times 10 ^9$ \cite{veverka1999near} & 2\\

 \hline
 \end{tabular}

  \caption{The number of potential impacts faced by a target body of a certain diameter. \\  }%
  \label{tab:noofimpacts}

\end{table}

\section{Conclusion}
\label{sec:concl}
In this work, we have presented a framework to model segregation on the surfaces of rubble asteroids over their lifetime due to impact-induced seismic shaking. For clarity of presentation, the model was developed and investigated in the context of an ellipsoidal rubble asteroid with the size and spin of Itokawa. However, the framework is easily adapted to any other asteroid. 

For this, we first estimated the seismic properties relevant to rubble bodies, modeled the manner in which seismic energy spreads through their interiors following an impact, and, finally, related the seismic energy received at a point to the surface vibrations there. Thus, for any given impact on a rubble body, we were able to estimate the amplitude and frequency of surface vibrations induced at any other location on that body. 

To follow the regolith's dynamical evolution over the lifetime of a near-Earth asteroid  (NEA), we need to know the number, frequency, size, and location of impacts that the NEA experiences. This is done stochastically. To this end, we first created several possible collisional histories by estimating the number of impacts that an NEA such as Itokawa may undergo over $10^6$ years and the size distribution of these impactors. Impacts could then take place at randomly distributed time intervals and surface locations. 


Finally, we set up discrete element simulations to numerically investigate the response of regolith on a rubble body to impact-induced seismic vibrations during its lifetime. In these simulations, we vertically vibrated a bin filled with a mixture of different sized grains every time an impact took place in any sample collisional history.  The results were then averaged over several collisional histories. The bin was located at the region of interest on the rubble body. The frequency and amplitude of the vibrations were related to the impact's magnitude and location, as mentioned above. The grain size distribution was taken as close as possible to the reported distribution for Itokawa at the bin's location. 



We then employed our framework to investigate segregation on our rubble ellipsoidal asteroid. We also probed the effect of various asteroidal properties on the segregation process. We found that seismic activity due to impacts is sufficient to drive size segregation on  small rubble asteroids. Medium and large boulders always rise to the top of the regolith. The end outcome is largely dependent upon the total energy supplied by large impact events in a collisional history, and not on the manner in which the impactor's size, frequency, and location were distributed. The degree of segregation, however, does vary with location on the asteroid. Most segregation is observed at centrally located regions and least at the most remote zones, i.e. segregation is best at the ends of the smallest principal axis of our ellipsoidal asteroid and worst at the ends of its longest axis. The quality of segregation is also affected by the wave speed and seismic diffusivity of the asteroid's interior, with low values of either leading to a smaller fraction  of medium and large boulders rising to the surface. These observations were explained on the basis of the travel times of the seismic waves and their dissipation in the asteroid's bulk.






Our model presents a systematic way to model seismic shaking on rubble asteroids and relate it to seismic activity and surface texturing.  Such a framework may be utilized to  explain the presence of large boulders on the surface of asteroids like Itokawa. Other asteroids, such as Bennu and Ryugu, may also be modeled with minor changes in the framework to account for their different geometry, rotation, and other parameters. Having said that, smooth regions like Muses C on Itokawa will require further investigation. Were only size segregation processes of the kind studied here active, then such regions cannot exist. We believe that explaining such observations  requires attention to the surface topography and the local gravity field, which need to be included within the present framework. This is  the focus of  ongoing work.

\enlargethispage{20pt}

\dataccess{The data that support the findings of this study are available from the corresponding author.} 

\competing{We declare we have no competing interests.}

\funding{S.G. would like to acknowledge the financial support during his M. Tech. from the Ministry for Education, Govt. of India,  when this work was done.}

\ack{The authors would like to thank the high-performance computing facility at the Indian Institute of Technology, Kanpur. }

\bibliographystyle{RS} 
\bibliography{Version_2023}
\end{document}